\documentclass[%
 reprint,
superscriptaddress,
nofootinbib,
 amsmath,amssymb,
 aps,
 prd,
]{revtex4-2}

\usepackage{graphicx}
\usepackage{dcolumn}
\usepackage{bm}
\RequirePackage[colorlinks=true
  ,urlcolor=blue
  ,anchorcolor=blue
  ,citecolor=blue
  ,filecolor=blue
  ,linkcolor=blue
  ,menucolor=blue
  ,linktocpage=true
]{hyperref}

\usepackage{url}
\usepackage{subfigure}
\usepackage{cellspace}
\begin{document}

\preprint{APS/123-QED}

\title{Reconstructing the dark energy density in light of DESI BAO observations}

\author{Maria Berti}
\email{maria.berti@unige.ch}
\affiliation{%
 Département de Physique Théorique and Center for Astroparticle Physics, Université de Genève, Quai E. Ansermet 24, CH-1211 Genève 4, Switzerland
}%
\author{Emilio Bellini}%
\affiliation{Center for Astrophysics and Cosmology, University of Nova Gorica, Nova Gorica, Slovenia}
\affiliation{INAF, Istituto Nazionale di Astrofisica, Osservatorio Astronomico di Trieste,\\ Via G. B. Tiepolo 11, I-34131 Trieste, Italy}
\affiliation{%
IFPU, Institute for Fundamental Physics of the Universe, via Beirut 2, I-34151 Trieste, Italy
}%

\author{Camille Bonvin}
\affiliation{%
 Département de Physique Théorique and Center for Astroparticle Physics, Université de Genève, Quai E. Ansermet 24, CH-1211 Genève 4, Switzerland
}%
\author{Martin Kunz}
\affiliation{%
 Département de Physique Théorique and Center for Astroparticle Physics, Université de Genève, Quai E. Ansermet 24, CH-1211 Genève 4, Switzerland
}%
\author{Matteo Viel}
\affiliation{%
 SISSA - International School for Advanced Studies, Via Bonomea 265, I-34136 Trieste, Italy
}%
\affiliation{%
INFN – National Institute for Nuclear Physics, Via Valerio 2, I-34127 Trieste, Italy
}%
\affiliation{%
IFPU, Institute for Fundamental Physics of the Universe, via Beirut 2, I-34151 Trieste, Italy
}%
\affiliation{INAF, Istituto Nazionale di Astrofisica, Osservatorio Astronomico di Trieste,\\ Via G. B. Tiepolo 11, I-34131 Trieste, Italy}
\affiliation{Italian Research Center on High Performance Computing, Big Data and Quantum Computing}
\author{Miguel Zumalacarregui}
\affiliation{Max Planck Institute for Gravitational Physics (Albert Einstein Institute), Am M\"{u}hlenberg 1, D-14476 Potsdam-Golm, Germany}

\date{\today}

\begin{abstract}
In light of the evidence for dynamical dark energy (DE) found from the most recent Dark Energy Spectroscopic Instrument (DESI) baryon acoustic oscillation (BAO) measurements, we perform a non-parametric, model-independent reconstruction of the DE density evolution. To do so, we develop and validate a new framework that reconstructs the DE density through a third-degree piece-wise polynomial interpolation, allowing for direct constraints on its redshift evolution without assuming any specific functional form. The strength of our approach resides in the choice of directly reconstructing the DE density, which provides a more straightforward relation to the distances measured by BAO than the equation of state parameter. We investigate the constraining power of cosmic microwave background (CMB) observations combined with supernovae (SNe) and BAO measurements. In agreement with results from other works, we find a preference for deviations from $\Lambda$CDM, with a significance of $2.4\sigma$ when using the Dark Energy Survey Year 5 (DESY5) SNe data, and $1.3\sigma$ with PantheonPlus. In all the cases we consider, the derived DE equation of state parameter presents evidence for phantom crossing. By investigating potential systematic effects in the low-redshift samples of DESY5 observations, we confirm that correcting for the offset in apparent magnitude with respect to PantheonPlus data, as suggested in previous studies, completely removes the tension. Furthermore, we assess the risk of potentially overfitting the data by changing the number of interpolation nodes. As expected, we find that with lesser nodes we get a smoother reconstructed behavior of the DE density, although with similar overall features. The pipeline developed in this work is ready to be used with future high-precision data to further investigate the evidence for a non-standard background evolution.
\end{abstract}

\maketitle


\section{Introduction}
It is a truth universally acknowledged that a modern theoretical cosmologist, in possession of evidence of a Universe dominated by dark components, must be in want of an efficient, unbiased way to test their model against observations. Although the existence of dark energy (DE) and dark matter (DM) seems indisputable, an equally clear understanding of their nature, the physics laws they obey, and their behavior in different scenarios is still lacking. Since the first observational evidences for cosmic acceleration \cite{SupernovaSearchTeam:1998fmf, SupernovaCosmologyProject:1998vns}, in the past decades a plethora of alternative scenarios {have been proposed}, ranging from the simplest variations of the DE equation of state to sophisticated modified gravity (MG) models (e.g.\ \citep{Ishak:2018his,Ferreira:2019xrr,Hou:2023kfp}).

A difficulty one encounters when facing the task of testing DE and gravity theories is the fact that it is not trivial to approach the problem in a complete model-independent, unbiased fashion. Even within the broad parametrized frameworks, like the phenomenological~\cite{Bertschinger:2006aw, Amendola:2007, Pogosian:2007sw, Zhao:2008, Zhang:2007} and the effective field theory (EFT)~\cite{Bloomfield:2012ff, Gubitosi:2012hu, Bellini:2014fua} descriptions, one still has to choose a specific time evolution for the functions that describe DE/MG effects. This, in turn, restricts the explored theory space, limiting the information that might be enclosed in observed data. Reconstruction techniques offer an alternative, agnostic approach to testing DE and MG. The rationale is to allow the measured data to directly drive the constraints on the time evolution of the targeted functions, without assuming any theory or prejudice-induced time dependence a priori~\cite{Huterer:2002hy}. One possibility is to use piece-wise parameterizations to constrain their value at given redshifts directly from observations. The method has been tested and applied in several scenarios, in particular to reconstruct the DE equation of state~\cite{Huterer:2002hy, Mortonson:2008qy, Zhao:2009fn,Zhao:2010dz, Mortonson_2010,Hojjati:2011xd, Zhao_2012, Zhao:2017cud, Casas:2017eob}. Recent results on the reconstructed DE background, phenomenological and EFT functions provided compelling insights on several open problems in cosmology~\cite{Raveri:2019mxg, Park:2021jmi, Pogosian:2021mcs, Raveri:2021dbu}. 

In particular, in the past months, the evidence for dynamical DE within the Chevallier-Polarski-Linder (CPL) parametrization found first with the Union3 supernovae (SNe) catalog~\cite{Rubin:2023ovl} and later combining the first baryon acoustic oscillation (BAO) measurements from the Dark Energy Spectroscopic Instrument (DESI)~\cite{DESI:2024cosmo_bao, DESI:2024uvr,DESI:2024full_shape} with cosmic microwave background (CMB) and SNe observations, further lighted the interest of the community into model-independent tests of DE. The $\sim 4\sigma$ tension found by the DESI collaboration has been extensively studied within several reconstruction scenarios, from expanding on a base of functions \cite{DESI:2024crossing} or using several parametric forms of the background \cite{DESI:2024kob}, to gaussian processes \cite{Mukherjee:2024ryz, Dinda:2024ktd,Cozzumbo:2024vxw,Yang:2025kgc}, reconstruction of the Hubble parameter \cite{Liu:2024gfy,LHuillier:2024rmp, Jiang:2024xnu}, and piece-wise reconstructions \cite{Pang:2024qyh, Ye:2024ywg}. At present, no conclusive evidence has been found to support one specific model, with thawing gravity being a possible interesting new candidate \cite{Ye:2024ywg, Ye:2024alone}. Indeed, the discussion among the scientific community is still open, waiting for {new data, such as the forthcoming} second {DESI data release}, to provide new information and shed light on the DE background.

Inspired by the methodology proposed in~\cite{Raveri:2019mxg,Pogosian:2021mcs} and the results from~\cite{Ye:2024ywg}, in this work, we perform a non-parametric, model-independent reconstruction of the DE density evolution to investigate further the possibility of a dynamical DE suggested in~\cite{DESI:2024cosmo_bao}. We aim to understand more deeply: $i)$ how the deviations from $\Lambda$CDM behave at different redshifts, $ii)$ what is driving the constraining power and the tension, $iii)$ how the chosen parametrization can impact the significance of the results. To this end, we reconstruct the evolution of the DE density by means of a third-degree polynomial interpolation. Following \cite{DESI:2024cosmo_bao}, we use DESI BAO observations combined with CMB data, from Planck 2018~\cite{planck:2018} and the Acatama Cosmology Telescope (ACT)~\cite{ACT:2023dou,ACT:2023kun} measurements, and the SNe catalogs from the Pantheon~\cite{Brout:2022vxf,Riess:2021jrx} and Dark Energy Survey (DES)~\cite{DES:2024tys,DES:2024hip} collaborations. Targeting directly the DE density, instead of the equation of state parameter, allows for a more straightforward interplay between the quantities measured by BAO and the reconstructed function. {The DE density is indeed more directly related to the distances measured with BAO than the equation of state, since the latter impacts the DE density through an integral over redshift. As a consequence, the DE density is ideally suited to capture a deviation from the expected value in a specific range of redshift.} We implement and test our pipeline within the Einstein-Boltzmann solver \texttt{class} \cite{Diego_Blas_2011} and conduct a Bayesian analysis to constrain the model parameters within the \texttt{cosmosis}~\cite{Zuntz:2014csq} framework.

The discussion is organized as follows. We describe our methodology in \autoref{sec:methods}, where we give an overview of the adopted reconstruction approach in \autoref{sec:recon}, the used observables in \autoref{sec:obs}, and the details of the statistical analysis in \autoref{sec:analysis_obs}. Results are discussed in \autoref{sec:results}. The full DE density reconstruction for the considered scenarios is presented in \autoref{sec:res_recon}. We attempt to disentangle the constraining power of the involved data sets in \autoref{sec:res_detangle}, also investigating the alleged systematic effect in the low redshift samples of DESY5 catalog. In \autoref{sec:res_nodes}, we discuss the impact of the number of nodes used to reconstruct the DE density, assessing the risk of overfitting the data. 
We provide a summary of the results and closing considerations in \autoref{sec:conclusions}.

\section{Methods}\label{sec:methods}
\subsection{Reconstructing the density evolution}\label{sec:recon}
We reconstruct the evolution of the background via a third-degree piece-wise polynomial interpolation. To better identify modifications from a cosmological constant contribution, we rewrite the density parameter of a generic DE fluid X as
\begin{equation}\label{eq:omega_def}
    \Omega_{\rm X}(z) = \Omega_{\Lambda} (z) \left[1 + \Delta\Omega_{\rm X}(z) \right],
\end{equation}
where $\Omega_{\rm X} (z) \equiv {\rho_{\rm X}(z)}/{\rho_{\rm cr}(z)} $ is the total fractional DE density and $\Omega_{\Lambda} (z) \equiv {\rho_\Lambda}/{\rho_{\rm cr}(z)} $ the contribution {that would be} given by a cosmological constant $\Lambda$.
We then reconstruct the evolution of the function $\Delta\Omega_{\rm X}(z)$, which encloses all deviations from a $\Lambda$CDM Universe. To ensure continuous and smooth first and second derivatives $\Delta\Omega_{\rm X}'(z)$ and $\Delta\Omega_{\rm X}''(z)$, the full evolution of $\Delta\Omega_{\rm X}(z)$ is obtained via a cubic spline interpolation among a fixed number $N$ of chosen nodes $\Delta\Omega^i_{\rm X}$ at the corresponding redshifts $z_i$, with $i=0, \dots N$. Moreover, we require $\Delta\Omega_{\rm X}(z)$ to assume a constant value $\overline{\Delta\Omega}_{\rm X}$ at high redshifts, i.e.\ we set $ \Delta\Omega_{\rm X}(z) = \overline{\Delta\Omega}_{\rm X}$ for $z\geq\bar{z}$. {This allows us to transition smoothly between the low redshift range where we have BAO and SNe measurements of distances, and the high redshift regime constrained by CMB.} 
Between the last node at $z_N$ and $\bar{z}$ we interpolate with an \textit{ad hoc} fifth-degree polynomial, imposing the smoothness and continuity of $\Delta\Omega_{\rm X}(z)$, its first and second derivatives in $z_N$ and setting $\Delta\Omega_{\rm X}'(\bar{z})=\Delta\Omega_{\rm X}''(\bar{z}) = 0$. 
The $N + 1$ nodes $\Delta\Omega^0_{\rm X}, \dots, \Delta\Omega^N_{\rm X}$ together with $\overline{\Delta\Omega}_{\rm X}$ can be introduced as new parameters in the analysis and constrained from observations.

We implement the non-parametric modeling of $\Omega_{\rm X}(z)$ presented above within the \texttt{\_fld} framework in the Einstein-Boltzmann solver \texttt{class}\footnote{See \url{http://class-code.net}.}~\cite{Diego_Blas_2011}. As originally proposed, this fluid includes a linearly varying (\textit{\`a la} CPL) equation of state $w(z)=p_X(z)/\rho_X(z)$ and a constant sound speed $c_s^2$ \cite{Ballesteros:2010ks}. In this work we fix $c_s^2=1$ and vary $\Delta\Omega_{\rm X}(z)$ instead of $w_{\rm X}(z)$. On top of this, one of the advantages of using the \texttt{\_fld} fluid is that it allows to cross the phantom divide, i.e.\ $w(z)=-1$, that would be forbidden in standard perfect fluid structures. To do so, it uses the parameterized post-Friedmann (PPF) approximation, as described in \cite{Fang:2008sn}.

While it is clear that this pipeline is data-inspired and has no exact mapping with specific theories, it can capture the phenomenology of single field DE, such as quintessence or k-essence. In addition, letting the phantom divide be crossed can give hints on deviations from simple DE, as suggested in \cite{DESI:2024cosmo_bao}.

From the reconstructed density, it is possible to derive the evolution of the equation of state parameter $w(z)$ for the general DE fluid. 
Imposing DE conservation, one finds that
\begin{equation}
     w(z) = \frac{1}{3} (1+z) \frac{1}{\rho_{\rm X}} \frac{d}{dz}{\rho_{\rm X}} -1.
\end{equation}
Using the definition $\Omega_{\rm X}(z) \equiv \rho_{\rm X}(z) /\rho_{\rm cr}(z)$
together with Eq.~(\ref{eq:omega_def}), the relation above translates to 
\begin{equation}\label{eq:w}
    w(z) = \frac{1+z}{3(1 + \Delta\Omega_{\rm X}(z))} \frac{d \Delta \Omega_{\rm X}(z)}{dz} -1. 
\end{equation}

We believe that our approach provides a more direct link with observations, relative to other works, in which the reconstruction is performed over $w(z)$. Indeed, {BAO observations measure $H(z)$ at a set of redshifts, which is directly related to $\Omega_{\rm X}(z)$ at the same redshifts} through the Friedmann equations. {On the contrary, an} additional integration step {is} required {to link $H(z)$ to} $w(z)$. {In addition, the angular diameter distance, measured through transverse BAO, and the luminosity distance, measured from SNe, are related to $\Omega_{\rm X}(z)$ through only one integral over redshift, whereas two integrals relate them to the equation of state. As a consequence, $\Omega_{\rm X}$ is better adapted to capture sharp changes in the observed distances.} 

\subsection{Observables}\label{sec:obs}
Since the very first evidence for cosmic acceleration \cite{SupernovaCosmologyProject:1998vns,SupernovaSearchTeam:1998fmf}, SNe observations and constraining DE have been inherently intertwined. In this work, we make use of several distance measures from SNe and BAO to constrain the evolution of the background in the quest to confirm the evidence for dynamical DE found with DESI observations \cite{DESI:2024uvr,DESI:2024cosmo_bao, DESI:2024full_shape}. In the following, we briefly define the observables we target in the analysis.

BAOs, usually referred to as standard rulers, provide a measure of distance in the directions parallel and transverse to the line of sight. For a flat Friedmann-Lemaître-Robertson-Walker (FLRW) metric, we can define the comoving distance as 
\begin{equation}
    D_M(z) \equiv \int_0^z \mathrm{d}z'\frac{c}{H(z')},
\end{equation}
where $H(z)$ is the Hubble parameter and $c$ the speed of light. {The comoving distance is related to the angular diameter distance by the redshift, and it can be directly measured from the transverse position of the BAO peak. The radial BAO peak provides instead a measurement of the distance} 
\begin{equation} 
D_H(z) \equiv \frac{c}{H(z)},
\end{equation}
{which directly depends on $H(z)$. Finally, by measuring the BAO peak averaged over all directions, from the monopole of the correlation function, one can measure} the volume distance $D_V$, {which is a} combination of the transverse and the radial distance
\begin{equation} D_V \equiv \left( z D_M(z)^2 D_H(z) \right)^{1/3}. 
\end{equation}
DESI observations, the key observable of this analysis, provided five measurements of $D_M$ and $D_H$, at the effective redshifts $z=0.51,\, 0.71, \, 0.93,\, 1.32,\, 2.33$, and two of $D_V$, at $z=0.29, \, 1.49$, {where the statistics is not good enough to allow for a separation into transverse and radial distances}. In our analysis, we include all the data points for the three observables and the seven effective redshifts, as described in the following section.

SNe are, instead, standardizable candles and effectively provide a measure of the luminosity distance $D_L(z)$, which is related to the comoving distance as
\begin{equation} D_M(z) = \frac{D_L(z)}{1+z}. \end{equation}
The apparent magnitude $m(z)$ of a SN relates to its luminosity distance as
\begin{equation} m(z) = 5\log_{10}[D_L(z)] + 25 + M, \end{equation}
where $M$ is the absolute magnitude of the object. The distance modulus $\mu(z)$ is then defined as:
\begin{equation} 
\mu(z) = m(z) - M .
\end{equation}

Following the approach of \cite{DESI:2024cosmo_bao}, we reconstruct the background evolution by combining BAO and SNe measurements with CMB data, as described in the following section.

\subsection{Analysis and used data sets}\label{sec:analysis_obs}
We set up the framework to constrain the cosmological parameters adopting a Bayesian approach. 

We reconstruct $\Delta\Omega_{\rm X}(z)$ using two different configurations. To isolate the constraining power from DESI BAO data at different redshifts, we decide to spline the function among eight nodes $\Delta\Omega^i_{\rm X}$, with $i=0, \dots 7$, at the observed DESI effective redshifts $z_{i} = \{0.0, 0.3, 0.51, 0.71, 0.93, 1.32, 1.49, 2.33\}$. In the following, we refer to this configuration as \textit{8-nodes case}. We highlight that we do not include any theory-informed prior on the shape of the reconstructed function. This allows for more freedom in exploring the parameter space at the cost, however, of risking overfitting the data (see e.g.\ \cite{Pogosian:2021mcs}). Thus, to investigate the impact of the chosen number of nodes on the constraining power, we also take into account a second configuration. We spline over four equispaced nodes in the redshift range of DESI observations, i.e.\ $z_{0,\ldots,3} = \{0.0, 0.8, 1.6, 2.4\}$ (\textit{4-nodes case}). It is well-known that the overfitting risk can be mitigated by introducing a correlation between the nodes. We leave the implementation of smoothing techniques studied in literature \cite{Crittenden:2011aa, Raveri:2019mxg} for future work.

For both cases, we anchor the evolution of the DE density at $z=0$, by fixing $\Delta\Omega^0_{\rm X} = 0$, while we choose to recover a constant, but not necessarily zero, $\Delta\Omega_{\rm X}(z)$ for $ z \geq \bar{z} = 4$, denoted as $\overline{\Delta\Omega}_{\rm X}$. It is possible to show that $\Delta\Omega^0_{\rm X}$ and $\overline{\Delta\Omega}_{\rm X}$ are completely degenerate. We choose to fix the DE fractional density today for a more standard interpretation of $\Omega_{\Lambda} (z=0)$.

We perform a Monte Carlo Markov Chain (MCMC) \cite{gilks:1995} analysis varying the six parameters describing the $\Lambda$CDM model, i.e.\ $\{\Omega_b h^2,\, \Omega_c h^2, n_s,\, \ln (10^{10} A_s),\, \tau,\, h \}$, together with the spline nodes $\Delta\Omega^i_{\rm X}$ (with $i = 1, \dots, 7$, $i=1, \dots 3$ for the 8, 4-nodes cases respectively) and the high redshift limit value $\overline{\Delta\Omega}_{\rm X}$. For all the parameters, we choose the uninformative, wide, flat priors given in \autoref{tab:prior}.

To efficiently sample the high-dimensional parameter space we use the affine invariant MCMC \cite{aff_inv} sampler \texttt{emcee}\footnote{See \url{https://emcee.readthedocs.io/en/stable}.} \cite{Foreman_Mackey_2013}, embedded in the \texttt{cosmoSIS} code standard library\footnote{See \url{https://cosmosis.readthedocs.io/en/latest}.}~\cite{Zuntz:2014csq}. To validate the output from \texttt{emcee}, we cross-check our results by running chains also with the Metropolis-Hastings algorithm~\cite{Metropolis:1953am,Hastings:1970aa} for selected cases.
The analysis of the MCMC samples to compute the marginalized constraints is performed with the Python package \texttt{GetDist}\footnote{See \url{https://getdist.readthedocs.io}.}~\citep{Lewis:2019xzd}. We also provide bestfit results obtained by means of Nelder-Meads minimization of the logarithmic likelihood \cite{Nelder:1965zz}.  
\begin{table}
\centering
\caption{Assumed flat priors on the model parameters.\label{tab:prior}}
\begin{ruledtabular}
\setlength{\extrarowheight}{2pt}
\begin{tabular}{lc}
Parameter  & Prior\\
		\hline
        $\Omega_b h^2$ & $[0.005, 0.1]$\\
        $\Omega_c h^2$  & $[0.001, 0.99]$\\
        $n_s$  & $[0.8, 1.2]$\\
        $\ln (10^{10} A_s)$  & $[1.61, 3.91]$\\
        $\tau$  & $[0.01, 0.8]$\\
        $h$ & $[0.2, 1]$\\ \hline
        $\Delta\Omega^i_{\rm X}$ & [-1,10] \\
        $\overline{\Delta\Omega}_{\rm X}$ & [-1,1000] \\
\end{tabular}
\end{ruledtabular}
\end{table}
\begin{figure}
\centering
\includegraphics[width=1\columnwidth]{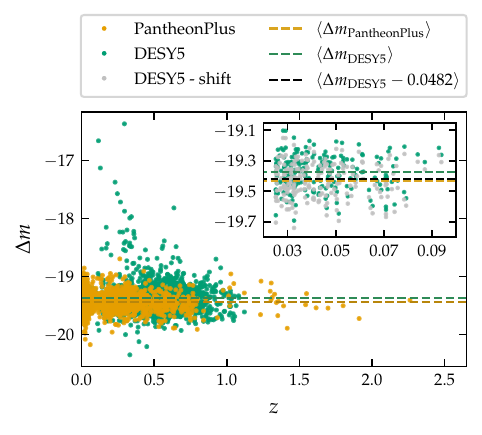}
\caption{\label{fig:mus} Observed magnitudes data points for PantheonPlus (yellow points) and DESY5 (green points). We show differences from the Planck 2018 predictions $\Delta m = m_{\rm observed} - m_{\rm Planck}$. To be consistent with~\cite{Efstathiou:2024xcq}, we convert the DESY5 observed $\mu(z)$ to $m(z) = \mu(z) -19.33$. The dashed lines correspond to the mean $\langle\Delta m\rangle$ of all measured data points in the two catalogs. We use the offset in magnitude found in~\cite{Efstathiou:2024xcq} (0.0482) to shift the samples at redshifts $z\leq 0.1$ in the DESY5 data set (gray points). We observe that after the shift on the low redshift data, the mean magnitudes of the two data sets are significantly more compatible. 
}
\end{figure}

We include in the analysis the following data sets and likelihoods:
\begin{itemize}
    \item BAO:  we use the five measurements of $D_{\rm M}$ and $D_{\rm H}$ and the two of $D_{\rm V}$ presented in Table 1 of \cite{DESI:2024cosmo_bao}. We implement a likelihood function for DESI observations in the \texttt{cosmoSIS} code. 
    In the following, we refer to this data set as ``DESI";
    \item CMB: we consider Planck PR3 observations~\citep{planck:2018}, including the high-$\ell$ TT, TE, EE from the \texttt{Plik lite} likelihood~\citep{planck:2018like} in the interval of multipoles $30\leq\ell\leq 2508$ for TT and $30\leq\ell\leq 19696$ for TE, EE. For the low-$\ell$ TT power spectrum, we use data from the \texttt{Commander} component-separation algorithm in the range $2 \leq \ell \leq 29$. We also adopt the Planck PR3 CMB lensing likelihood and the low EE polarization power spectrum in the range $2 \leq \ell\leq 29$, calculated from the likelihood code \texttt{SimAll}~\citep{planck:2018_maps}. For lensing measurements, we also include the latest Acatama Cosmology Telescope (ACT) DR6 data \cite{ACT:2023dou,ACT:2023kun}. In the following, we refer to the combination of all CMB datasets listed here with the label ``CMB";
    \item SNe: we consider two datasets, i.e.\ the 1550 SN Type IA (SNIA) from the Pantheon+ sample~\cite{Brout:2022vxf,Riess:2021jrx} in the redshift range $0.001 < z <2.26$ (``PantheonPlus"), and the 1829 SNIA in the Dark Energy Survey (DES) Year 5 sample~\cite{DES:2024tys,DES:2024hip}, spanning the redshift range $0.025 < z <1.3$ (``DESY5"). For both datasets we use the likelihood functions implemented in \texttt{cosmoSIS}.\footnote{See \url{https://github.com/PantheonPlusSH0ES} for PantheonPlus and \url{https://github.com/des-science/DES-SN5YR} for DESY5.} When using SNIA data sents, we marginalize over the absolute magnitude $M$. 
\end{itemize}
As in \cite{DESI:2024cosmo_bao}, we constrain the background evolution for different combinations of the datasets presented above. 

\subsubsection{A test on the impact of DESY5 low redshift samples}\label{sec:desy5_shift}
The robustness of the evidence for dynamical DE found with DESI data has been a prime interest of the scientific community. The discussion, prompted mainly by the results from \cite{Efstathiou:2024xcq}, is focused on understanding if the tension with the standard model is driven by truly new physics or by not-well-understood systematic effects. In particular, in \cite{Efstathiou:2024xcq} the author raises the concern that the low redshift samples in the DESY5 data set might be affected by a systematic that causes an offset in the observed magnitude with respect to PantehonPlus. Correcting for the shift seems to reduce the tension significantly, bringing the constraints on the background evolution back to $\Lambda$CDM.

Although the DES collaboration reasserted their results, excluding the possibility of overlooked systematics effects \cite{DES:2025tir}, the debate among the community is still lively \cite{Dhawan:2024gqy,Notari:2024zmi,Colgain:2024mtg, Huang:2025som}. 

Intrigued by the discussion, we test the argument from \cite{Efstathiou:2024xcq} also within our framework. From Table 1 in \cite{Efstathiou:2024xcq}, the low redshifts ($z\leq 0.1$) samples in the DESY5 catalog show an offset in the apparent magnitude of $-0.0482 \pm 0.0057$. We use this {central} value to shift the data at $z\leq 0.1$ in the DESY5 data set, leaving the covariance matrix unchanged, and investigate how the constraining power on the DE density changes compared with the official DESY5 observations. In the following, we refer to results from this test with the label ``DESY5 - shift". 

A graphical representation of our procedure is shown in \autoref{fig:mus}. We plot the magnitude difference with respect to a Planck 2018 fiducial magnitude together with the mean magnitude for both DESY and PantheonPlus. We observe that correcting for the measured offset indeed makes the two data sets more compatible.
\section{Results}\label{sec:results}
\subsection{Reconstructed dark energy density}\label{sec:res_recon}
\begin{figure*}
\includegraphics{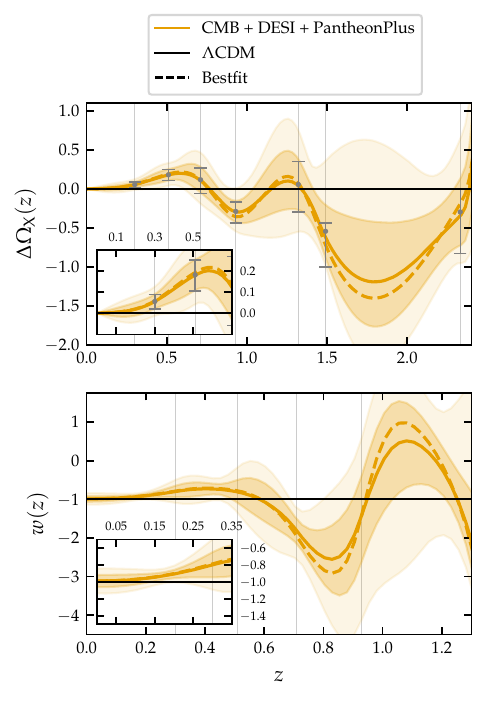}\includegraphics{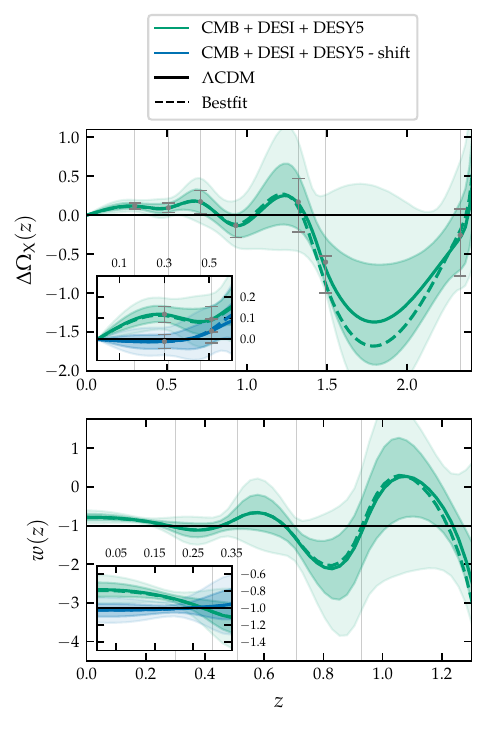}
\caption{\label{fig:shift_ev} Reconstructed background from CMB + DESI data combined with PantheonPlus SNe observations (\textit{left}) and DESY5 (\textit{right}). We show the reconstructed DE density deviations from $\Lambda$CDM $\Delta\Omega_{\mathrm{X}}(z)$, defined in Eq.~\ref{eq:omega_def} (\textit{upper panels}), and the derived evolution for the equation of state parameter $w(z)$ (\textit{lower panels}). Continuous solid lines correspond to the 50$^{\text{th}}$ percentile (mean) of all the trajectories explored in the MCMC. Dashed lines correspond to the evolution from bestfit results. Coloured and shaded areas represent the $1$ and $2\, \sigma$ regions, i.e.\ the 19$^{\text{th}}$ - 84$^{\text{th}}$ and 2$^{\text{th}}$ - 98$^{\text{th}}$ percentile ranges among all the explored trajectories respectively.
Grey data points represent the marginalized constraints on the single nodes, presented in \autoref{tab:constraints}. The solid black line marks the $\Lambda$CDM limit. Vertical lines correspond to the redshifts of the nodes chosen for the interpolation.
}
\end{figure*}

\begin{table*}
\caption{Marginalized constraints on the cosmological parameters at the $68\%$ confidence level. For upper and lower limits, $95\%$ confidence levels are given. Differences in $\chi^2$ are computed with respect to $\Lambda$CDM constraints obtained for the same combination of data sets. We highlight that the parameter $\overline{\Delta\Omega}_{\rm X}$ results to be unconstrained in all the considered scenarios. The reconstructed evolution and confidence regions for the same data sets are shown in Figures \ref{fig:shift_ev} to \ref{fig:contour_cosmological_shift}.\label{tab:constraints} 
}
\setlength{\extrarowheight}{2pt}
\begin{ruledtabular}
\begin{tabular}{cccc}
{\textsc{Parameter}}  & CMB + DESI + PantheonPlus & CMB + DESI + DESY5 & CMB + DESI + DESY5 - shift \\ \hline
$h $  & $ 0.6816\pm 0.0075$ & $ 0.6722\pm 0.0069$ & $ 0.6904\pm 0.0071$  \\
$\Omega_ch^2 $  & $ 0.12029\pm 0.00091$ & $ 0.12029\pm 0.00089$ & $ 0.12027\pm 0.00093$  \\
$\Omega_bh^2 $  & $ 0.02235\pm 0.00013$ & $ 0.02235\pm 0.00014$ & $ 0.02235\pm 0.00014$  \\
$\tau $  & $ 0.0565\pm 0.0073$ & $ 0.0567\pm 0.0072$ & $ 0.0570^{+0.0070}_{-0.0078}$  \\
$n_s $  & $ 0.9649\pm 0.0036$ & $ 0.9649\pm 0.0036$ & $ 0.9649\pm 0.0036$  \\
$A_s \times 10^{9} $  & $ 2.111\pm 0.029$ & $2.111^{+0.027}_{-0.030}$ & $ 2.113^{+0.028}_{-0.031}$  \\ \hline
$\Delta\Omega^7_\mathrm{X} $  & $ -0.29^{+0.30}_{-0.54}$ & $ -0.26^{+0.33}_{-0.53}$ & $ -0.33^{+0.30}_{-0.48}$  \\
$\Delta\Omega^6_\mathrm{X}$  & $<0.137$ & $<0.0593$ & $<0.0185$  \\
$\Delta\Omega^5_\mathrm{X} $  & $ 0.06^{+0.30}_{-0.35}$ & $ 0.17^{+0.30}_{-0.39}$ & $ 0.06^{+0.31}_{-0.35}$  \\
$\Delta\Omega^4_\mathrm{X} $  & $ -0.29^{+0.12}_{-0.15}$ & $ -0.13^{+0.14}_{-0.16}$ & $ -0.19\pm 0.14$  \\
$\Delta\Omega^3_\mathrm{X} $  & $ 0.12^{+0.15}_{-0.18}$ & $ 0.17^{+0.14}_{-0.16}$ & $ 0.08^{+0.13}_{-0.14}$  \\
$\Delta\Omega^2_\mathrm{X} $  & $ 0.182\pm 0.073$ & $ 0.096\pm 0.060$ & $ 0.040\pm 0.057$  \\
$\Delta\Omega^1_\mathrm{X} $  & $ 0.055\pm 0.035$ & $ 0.119\pm 0.038$ & $ -0.0096\pm 0.033$  \\
\hline
$\Omega_m $  & $ 0.3085\pm 0.0071$ & $ 0.3172\pm 0.0068$ & $ 0.3007\pm 0.0065$  \\
$\sigma_8 $  & $ 0.8237\pm 0.0086$ & $ 0.8159\pm 0.0083$ & $ 0.8311\pm 0.0082$  \\\hline
$\Delta\chi^2$   & 11.35 & 18.84 & 1.41\\
Tension & 1.33$\sigma$ & 2.42$\sigma$ & - \\
\end{tabular}
\end{ruledtabular}
\end{table*}

{We start by showing the results} on {$\Delta\Omega_{\rm X}(z)$} for the 8-nodes case. Given the high number of parameters in this configuration, we find that CMB and BAO observations, both alone and combined, do not hold enough constraining power, leaving $\Delta\Omega_{\rm X}(z)$ completely unconstrained. We thus show results only in combination with the SNe measurements. We present the reconstructed background evolution from CMB + DESI + PantheonPlus and CMB + DESI + DESY5 in \autoref{fig:shift_ev} and \autoref{tab:constraints}. 

Our results are qualitatively in agreement with the findings in~\cite{DESI:2024cosmo_bao}, with $\Delta\Omega_{\rm X}(z)$ showing a preference for beyond-$\Lambda$CDM behavior at low redshifts, particularly below $z=1$. As we discuss in more detail in the next section, both with PantheonPlus and DESY5 there is at least one node {that is} more than $2\sigma$ away from the $\Lambda$CDM limit ($\Delta\Omega_{\rm X}^i = 0$). At higher redshifts, instead, $\Delta\Omega_{\rm X}(z)$ becomes more evidently compatible with $\Lambda$CDM within $2\sigma$ for both data sets. Within our framework, the overall significance of the deviations is reduced with respect to results in \cite{DESI:2024cosmo_bao}. Indeed, the reconstructed evolution is preferred with respect to $\Lambda$CDM at the 1.3$\sigma$ level, when using PantheonPlus, and 2.4$\sigma$ with DESY5, to be compared with the {respective} 2.5$\sigma$ {and} 3.9$\sigma$ values {obtained} with the CPL parametrization in \cite{DESI:2024cosmo_bao}. 

{On the top panels of \autoref{fig:shift_ev} it is possible to notice trajectories with $\Delta\Omega_X(z)<-1$, for redshifts $1.5<z<2.5$. This can happen between the nodes, where no prior has been assumed (while at the nodes, we impose $\Delta\Omega_X(z)\geq-1$, see \autoref{tab:prior}). From \autoref{eq:omega_def}, this implies a negative DE density at those times. In this context, a slightly negative DE density is not worrisome, indeed: \textit{i)} our framework is purely phenomenological, \textit{ii)} at those redshifts ($z\simeq2$) DE is subdominant ($\Omega_m/\Omega_\Lambda\sim 10^{-1}$) which results in a} negligible modification of the
{expansion history, and \textit{iii)}} the effect is induced by the use of the spline and 
{vanishes with fewer nodes (see the \textit{4-nodes} case in \autoref{sec:res_nodes}). If one wants to take this result more seriously, it could be interpreted as a preference of the data for MG models that can be described by} a
{negative density, e.g.~Eq.~(16) of \cite{Zumalacarregui:2020cjh} where an effective Planck mass smaller than the bare one can have exactly this effect.}

The low redshift differences between PantheonPlus and DESY5 propagate also in the equation of state. In the lower panels of \autoref{fig:shift_ev}, we show the evolution of the equation of state parameter $w(z)$, derived from $\Delta\Omega_{\rm X}(z)$, as discussed in \autoref{sec:recon}. We stress that beyond $z=1.2$ the error bars on $w(z)$ increase significantly. For this reason, we do not show the higher redshift behaviour. We observe that $w(z)$ is broadly compatible with a cosmological constant contribution. However, in both cases, it seems that the data lead to a crossing of the phantom divide, particularly with PantheonPlus, in agreement with other results in the literature. 

Indeed, several works investigated the reconstruction of $w(z)$ with different techniques using the same set of observations (see e.g.~\cite{DESI:2024crossing,Ye:2024ywg}). Given the similarity in the adopted techniques, we compare the results presented in this work with the $w(z)$ reconstructed in \cite{Ye:2024ywg} from CMB + DESI + PantheonPlus data. The two methodologies differ in: \textit{i)} the number of nodes and the redshift range (but the number of nodes is the same for $z \leq 1$), \textit{ii)} the direct (in \cite{Ye:2024ywg}) or derived (this work) reconstruction of $w(z)$, \textit{iii)} the Einstein-Boltzmann solver used.\footnote{{It should be noted that, even formally different, the approach used in \cite{Ye:2024ywg} to fix DE at early times is equivalent to ours. Their strategy consists in fixing $w(z>1)=-1$ and choosing standard initial conditions for the DE density (thus obtaining $\Lambda$CDM at early times), while we fix $\Delta\Omega_{\rm X}(z=0)=0$ and let the high redshift $\overline{\Delta\Omega}_{\rm X}$ to vary. Our approach still reproduces a $\Lambda$CDM early-times evolution with $\rho_{\rm X}=\rho_\Lambda\left(1+\overline{\Delta\Omega}_{\rm X}\right)$, and what we identified with $\rho_\Lambda$ in \autoref{eq:omega_def} is the DE density today.}}
Comparing the lower, left panel of \autoref{fig:shift_ev} with Figure 2 in \cite{Ye:2024ywg}, we observe that the reconstructed $w(z)$ in the two cases share comparable features. The only substantial difference is noticeable around $z=0$, where results from this work show a $w(z)$ more compatible with a cosmological constant. This comparison serves as a validation test of our new pipeline compared with already widely tested numerical tools.
{It is worth commenting that, as in \cite{DESI:2024cosmo_bao}, our results show a preference for models that cross the phantom divide. While this could be a statistical fluke, it may be an indication of DE/MG models beyond simple quintessence \cite{Chudaykin:2024gol}. Indeed, it is well known that a single scalar field \textit{\`a la} quintessence can not cross the phantom divide. Our framework is purely phenomenological, and by construction, it can not point to any specific model, but it can indicate promising directions, such as dark energy models involving multiple fields or MG. However, a proper analysis of those models on the same data should be performed before drawing any conclusion.}

We notice that both $\Delta\Omega_{\rm X}(z)$ and $w(z)$ manifest an oscillating behavior. This is a well-known feature of the interpolation method we implemented~\cite{Pogosian:2021mcs}. We refer to \autoref{sec:res_nodes} for a discussion on the potential risk of overfitting.

In the following section, we compare in more detail the constraining power of the DESY5 and PantheonPlus data sets, analyzing their features at different redshifts.

\begin{table*}
\caption{Marginalized constraints on the cosmological parameters at the $68\%$ confidence level. For upper and lower limits $95\%$ confidence levels are given. Differences in $\chi^2$ are computed with respect to $\Lambda$CDM constraints obtained for the same combination of data sets. We highlight that the parameter $\overline{\Delta\Omega}_{\rm X}$ results to be unconstrained in all the considered scenarios. The reconstructed evolution and confidence regions for the same data sets are shown in Figures \ref{fig:contour} to \ref{fig:contour_cosmological_shift}.\label{tab:constraints_no_DESI} 
}
\setlength{\extrarowheight}{2pt}
\begin{ruledtabular}
\begin{tabular}{cccc}
{\textsc{Parameter}}  & CMB + PantheonPlus & CMB + DESY5 & CMB + DESY5 - shift \\ \hline
$h $  & $ 0.675\pm 0.016$ & $ 0.629^{+0.015}_{-0.020}$ & $ 0.647\pm 0.016$  \\
$\Omega_ch^2 $  & $ 0.1207\pm 0.0011$ & $ 0.1213^{+0.0013}_{-0.0011}$ & $ 0.1212\pm 0.0012$  \\
$\Omega_bh^2 $  & $ 0.02232\pm 0.00014$ & $ 0.02228\pm 0.00015$ & $ 0.02228\pm 0.00015$  \\
$\tau $  & $ 0.0564^{+0.0070}_{-0.0079}$ & $ 0.0573\pm 0.0073$ & $ 0.0571\pm 0.0075$  \\
$n_s $  & $ 0.9637\pm 0.0039$ & $ 0.9627\pm 0.0042$ & $ 0.9628\pm 0.0042$  \\
$A_s \times 10^{9} $  & $ 2.113^{+0.028}_{-0.033}$ & $ ,2.120\pm 0.030$ & $ 2.119\pm 0.030$  \\ \hline
$\Delta\Omega^7_\mathrm{X}$  & $<1.56$ & $<2.11$ & $<1.59$  \\
$\Delta\Omega^6_\mathrm{X}$  & $<3.37$ & $<3.48$ & $<3.50$  \\
$\Delta\Omega^5_\mathrm{X}$  & $<2.25$ & $<5.42$ & $<3.73$  \\
$\Delta\Omega^4_\mathrm{X} $  & $<0.0464$ & $ 2.2^{+1.0}_{-1.2}$ & $ 2.04^{+0.90}_{-1.2}$  \\
$\Delta\Omega^3_\mathrm{X} $  & $ 0.16^{+0.33}_{-0.41}$ & $ -0.11^{+0.25}_{-0.28}$ & $ -0.25\pm 0.23$  \\
$\Delta\Omega^2_\mathrm{X} $  & $ 0.15\pm 0.11$ & $ -0.17\pm 0.11$ & $ -0.202^{+0.11}_{-0.090}$  \\
$\Delta\Omega^1_\mathrm{X} $  & $ 0.041^{+0.047}_{-0.042}$ & $ 0.067^{+0.059}_{-0.048}$ & $ -0.068^{+0.048}_{-0.041}$  \\ \hline
$\Omega_m $  & $ 0.316^{+0.015}_{-0.017}$ & $ 0.365^{+0.024}_{-0.021}$ & $ 0.345\pm 0.019$  \\
$\sigma_8 $  & $ 0.819\pm 0.015$ & $ 0.779^{+0.015}_{-0.018}$ & $ 0.793\pm 0.015$  \\
\hline
$\Delta\chi^2$ & 7.09 & 14.14 & 1.41\\
Tension & 0.63$\sigma$ & 1.76$\sigma$ & 0.32$\sigma$ \\
\end{tabular}
\end{ruledtabular}
\end{table*}

\begin{figure}
\centering
\includegraphics[width=1\columnwidth]{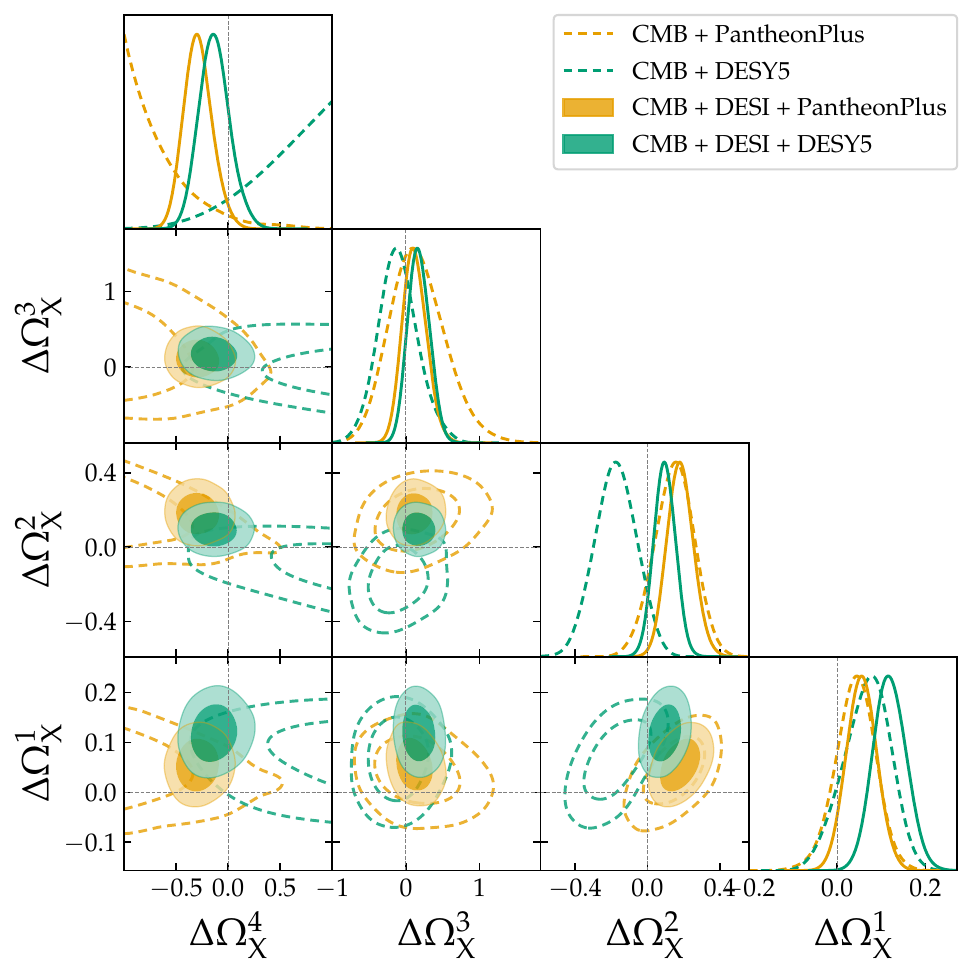}
\caption{\label{fig:contour} Joint constraints (68\% and 95\% confidence regions) and marginalized posterior distributions on the lowest redshift nodes of $\Delta\Omega_{\rm X}(z)$ (Eq.~\ref{eq:omega_def}), corresponding to the redshifts $z_{1,2,3,4} = \{0.3, 0.51, 0.71, 0.93\}$. Gray dotted lines mark the $\Lambda$CDM limit.}
\end{figure}

\begin{figure}
\centering
\includegraphics[width=1\columnwidth]{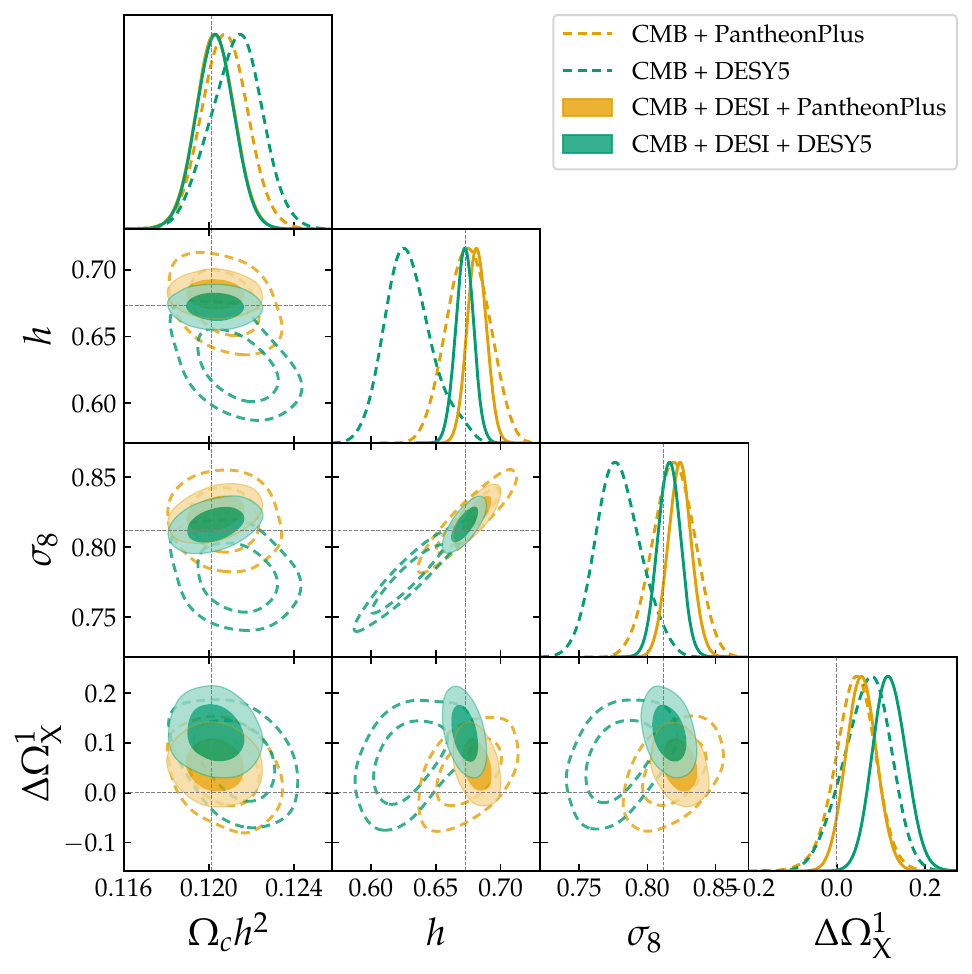}
\caption{\label{fig:contour_cosmological} Joint constraints (68\% and 95\% confidence regions) and marginalized posterior distributions on a subset of the cosmological parameters and the lowest redshift node of $\Delta\Omega_{\rm X}(z)$ (Eq.~\ref{eq:omega_def}), corresponding to the redshift $z_{1} = 0.3$. Gray dotted lines mark the Planck 2018 cosmology and the $\Lambda$CDM fiducial for $\Delta\Omega_{\rm X}^1$.
}
\end{figure}
\begin{figure}
\centering
\includegraphics[width=1\columnwidth]{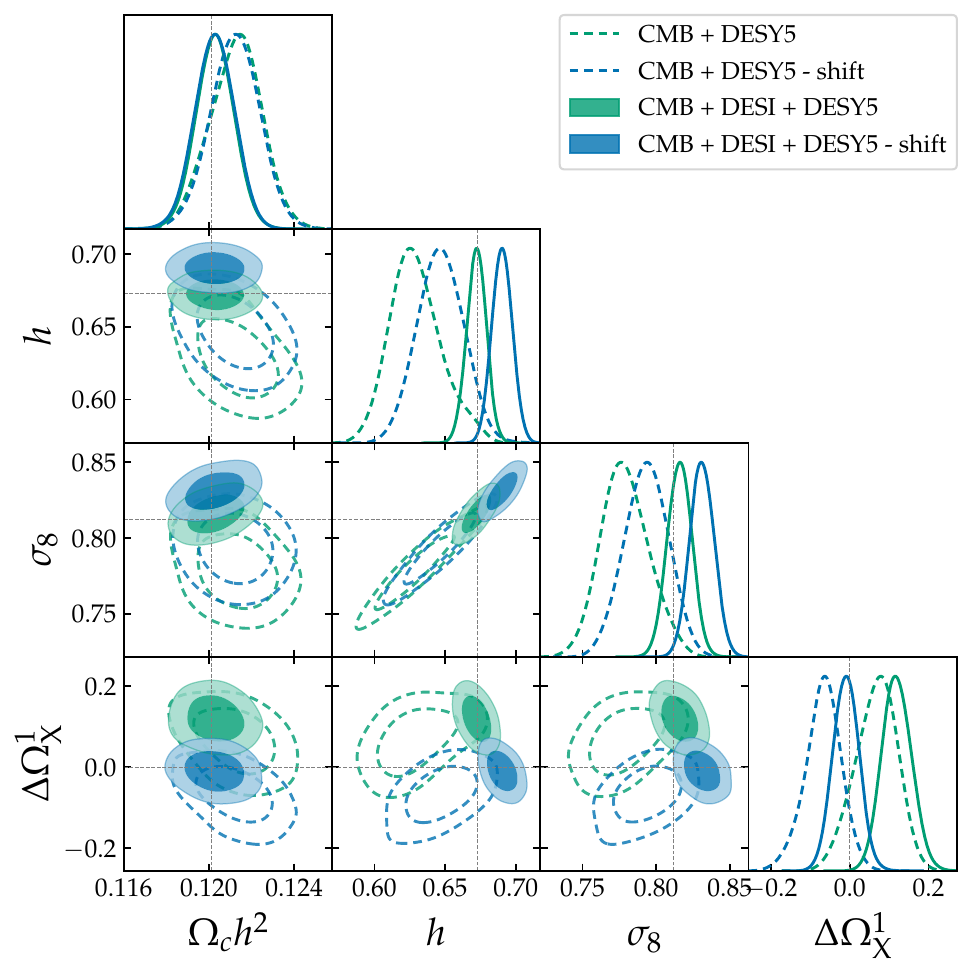}
\caption{\label{fig:contour_cosmological_shift} Joint constraints (68\% and 95\% confidence regions) and marginalized posterior distributions on a subset of the cosmological parameters and the lowest redshift node of $\Delta\Omega_{\rm X}(z)$ (Eq.~\ref{eq:omega_def}), corresponding to the redshift $z_{1} = 0.3$. Gray dotted lines mark the Planck 2018 cosmology and the $\Lambda$CDM fiducial for $\Delta\Omega_{\rm X}^1$.}
\end{figure}

\subsection{Disentangling constraining powers}\label{sec:res_detangle}
From the results presented in \cite{DESI:2024cosmo_bao}, it seems clear that only the specific combination of CMB, BAO DESI, and SNe observations show a preference for dynamical DE. The tension with a $\Lambda$CDM background is strongest when using the DESY5 catalog, leading to the conclusion that it is the combination of DESY5 and DESI BAO that actually drives the constraints away. In this section, we exploit the potential of our approach to appreciate features at different redshifts in order to investigate more deeply the constraining power of the involved data sets. In addition to the previous section, we discuss also the constraining power of CMB data combined only with SNe and the results from the test on DESY5 low redshift data points, described in \autoref{sec:desy5_shift}.

From \autoref{fig:contour_cosmological}, \autoref{fig:contour}, and \autoref{tab:constraints_no_DESI}, the differences between the CMB + PantheonPlus and CMB + DESY5 effects on the cosmological parameters can be appreciated. We notice that PantheonPlus and DESY5 prefer different values of $h$, with PantheonPlus being more compatible with a Planck {2018} cosmology. Both data sets, instead, agree on the other cosmological parameters. Also, in the constraints on $\Delta\Omega_{\rm X}(z)$ PantheonPlus provides results broadly more in agreement with $\Lambda$CDM than DESY5. It is interesting to see how, on some of the nodes, the two datasets seem to push the parameters in almost opposite directions, e.g.\ on $\Delta\Omega_{\rm X}^2$. When using CMB + DESY5 we find a mild tension with $\Lambda$CDM at the level of $1.76\sigma$, while CMB + PantheonPlus data are more in agreement with $\Lambda$CDM ($0.63\sigma$).

Adding DESI observations significantly lowers the estimated errors, reducing the degeneracies on most of the parameters, thus effectively increasing the tension. DESI data have a strong effect in particular when combined with DESY5. In particular, DESI has the effect of pushing back to an $h$ value closer to the one preferred by PantheonPlus. For the background, instead, we can see that although DESY5 is more in tension with $\Lambda$CDM, the effect seems to be completely driven only by the constraints on the lowest redshift node. Indeed, in the first redshift bin ($\Delta\Omega_{\rm X}^1$), DESY5 pushes the constraints away from $\Lambda$CDM more than PantheonPlus. However, in the second and fourth ($\Delta\Omega_{\rm X}^2$ and $\Delta\Omega_{\rm X}^4$), this behavior is inverted. We conclude that, although the significance of the deviation from $\Lambda$CDM is higher for the CMB + DESI + DESY5 case, the tension is driven by the constraint on $\Omega_{\rm X}^1$, at redshift $z=0.3$, while with PantheonPlus we observe a mild discrepancy with $\Lambda$CDM also at redshift $z=0.93$.

Shifting the low data points in the DESY5 sample leads to interesting results. From \autoref{fig:contour_cosmological_shift}, it is evident that the background evolution is significantly more in agreement with a cosmological constant contribution for the case CMB + DESI + DESY5-shift. Not only are the constraints on $\Delta\Omega_{\rm X}(z)$ generally more compatible with results from PantheonPlus, as found also in \cite{Efstathiou:2024xcq}, but the tension is also completely removed. The effect of the shift is the strongest on the lowest redshift node, as also clearly visible in the upper right panel of \autoref{fig:shift_ev}. By moving only few points in the data set, the constraint on $\Delta\Omega_{\rm X}^1$ is lowered from $0.119 \pm 0.038$ to $-0.0096\pm0.033$. Moreover, the derived $w(z)$ evolution results are much more compatible with the PantheonPlus case and a $\Lambda$CDM evolution. We notice, however, that the $\Lambda$CDM limit is recovered on the background at the expense of a shift in the cosmological parameters, such as on $h$ that is now pushed towards values even higher than Planck 2018. Thus, correcting for the offset in the DESY5 low redshift samples removes the tension with respect to a $\Lambda$CDM evolution, however preferring a cosmology less in agreement with Planck 2018. 

Interestingly, without DESI observations the results from the test are not equally clear. Shifting the low redshift points in DESY5 does not improve the agreement between CMB + DESY5 and CMB + PantheonPlus. Contrary to the case including DESI, results on $\Delta\Omega_{\rm X}^1$ are not more clearly in agreement with $\Lambda$CDM. On the other hand, the cosmological parameters are more in agreement with Planck 2018.

Although the test we perform serves only as a proof of concept, it seems to be yet another indication of a possible misunderstood effect, amplified when combining DESY5 with DESI observations. We leave for future work to investigate this issue further with the next data release from DESI BAO observations.

\subsection{Impact of the number of used nodes}\label{sec:res_nodes}
\begin{figure*}
\includegraphics{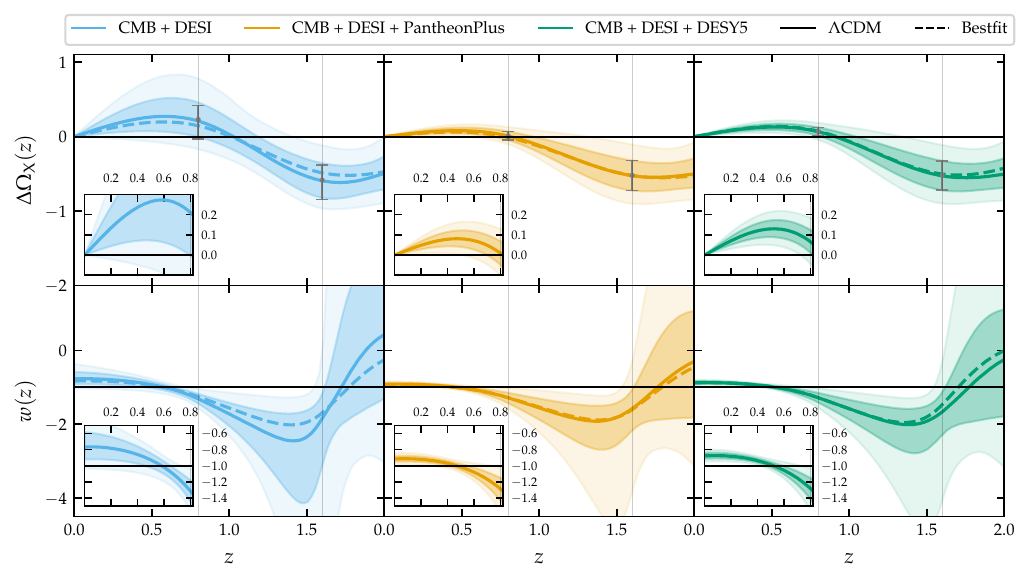}
\caption{\label{fig:shift_ev_3} Results for the 4-nodes case: reconstructed background from CMB + DESI data combined with PantheonPlus SNe observations (\textit{left}) and DESY5 (\textit{right}). We show the reconstructed DE density deviations from $\Lambda$CDM $\Delta\Omega_{\mathrm{X}}(z)$, defined in Eq.~\ref{eq:omega_def} (\textit{upper panels}), and the derived evolution for the equation of state parameter $w(z)$ (\textit{lower panels}). Continuous solid lines correspond to the 50$^{\text{th}}$ percentile (mean) of all the trajectories explored in the MCMC. Dashed lines correspond to the evolution from bestfit results. Coloured and shaded areas represent the $1$ and $2\, \sigma$ regions, i.e.\ the 19$^{\text{th}}$ - 84$^{\text{th}}$ and 2$^{\text{th}}$ - 98$^{\text{th}}$ percentile ranges among all the explored trajectories respectively. 
Grey data points represent the marginalized constraints on the single nodes, presented in \autoref{tab:constraints}. The solid black line marks the $\Lambda$CDM limit. Vertical lines correspond to the redshifts of the nodes chosen for the interpolation. 
}
\end{figure*}
\begin{table*}
\caption{Marginalized constraints on the cosmological parameters at the $68\%$ confidence level. For upper and lower limits $95\%$ confidence levels are given. Differences in $\chi^2$ are computed with respect to $\Lambda$CDM constraints obtained for the same combination of data sets. We highlight that the parameter $\overline{\Delta\Omega}_{\rm X}$ results to be unconstrained in all the considered scenarios. The reconstructed evolution and confidence regions for the same data sets are shown in \autoref{fig:shift_ev_3} and \ref{fig:contour_3}.\label{tab:constraints_3}}
\setlength{\extrarowheight}{2pt}
\begin{ruledtabular}
\begin{tabular}{cccc}
{\textsc{Parameter}}  & CMB + DESI & CMB + DESI + PantheonPlus & CMB + DESI + DESY5  \\ \hline
$h $  & $ 0.658\pm 0.024$ & $ 0.6812\pm 0.0072$ & $ 0.6738\pm 0.0067$  \\
$\Omega_ch^2 $  & $ 0.12043\pm 0.00092$ & $ 0.12031\pm 0.00093$ & $ 0.12039\pm 0.00092$  \\
$\Omega_bh^2 $  & $ 0.02235\pm 0.00014$ & $ 0.02235\pm 0.00013$ & $ 0.02235\pm 0.00014$  \\
$\tau $  & $ 0.0563\pm 0.0074$ & $ 0.0567^{+0.0069}_{-0.0078}$ & $ 0.0566^{+0.0069}_{-0.0078}$  \\
$n_s $  & $ 0.9646\pm 0.0036$ & $ 0.9647\pm 0.0036$ & $ 0.9645\pm 0.0036$  \\
$A_s \times 10^{9} $  & $ 2.110\pm 0.030$ & $ 2.112^{+0.027}_{-0.031}$ & $ 2.111^{+0.027}_{-0.031}$  \\ \hline
$\Delta\Omega^3_\mathrm{X} $  & $ -0.06^{+0.37}_{-0.71}$ & $ -0.27^{+0.32}_{-0.55}$ & $ -0.26^{+0.31}_{-0.58}$  \\
$\Delta\Omega^2_\mathrm{X} $  & $ -0.58^{+0.20}_{-0.26}$ & $ -0.52\pm 0.20$ & $ -0.52\pm 0.20$  \\
$\Delta\Omega^1_\mathrm{X} $  & $ 0.23^{+0.19}_{-0.26}$ & $ 0.010\pm 0.059$ & $ 0.067\pm 0.057$  \\ \hline
$\Omega_m $  & $ 0.333^{+0.022}_{-0.027}$ & $ 0.3090\pm 0.0068$ & $ 0.3159\pm 0.0066$  \\
$\sigma_8 $  & $ 0.803\pm 0.022$ & $ 0.8234\pm 0.0086$ & $ 0.8176\pm 0.0081$  \\ \hline
$\Delta\chi^2$ & 6.27 & 4.5 & 13.87\\
Tension & 1.34$\sigma$ & 0.98$\sigma$ & 2.66$\sigma$\\
\end{tabular}
\end{ruledtabular}
\end{table*}
\begin{figure}
\centering
\includegraphics[width=1\columnwidth]{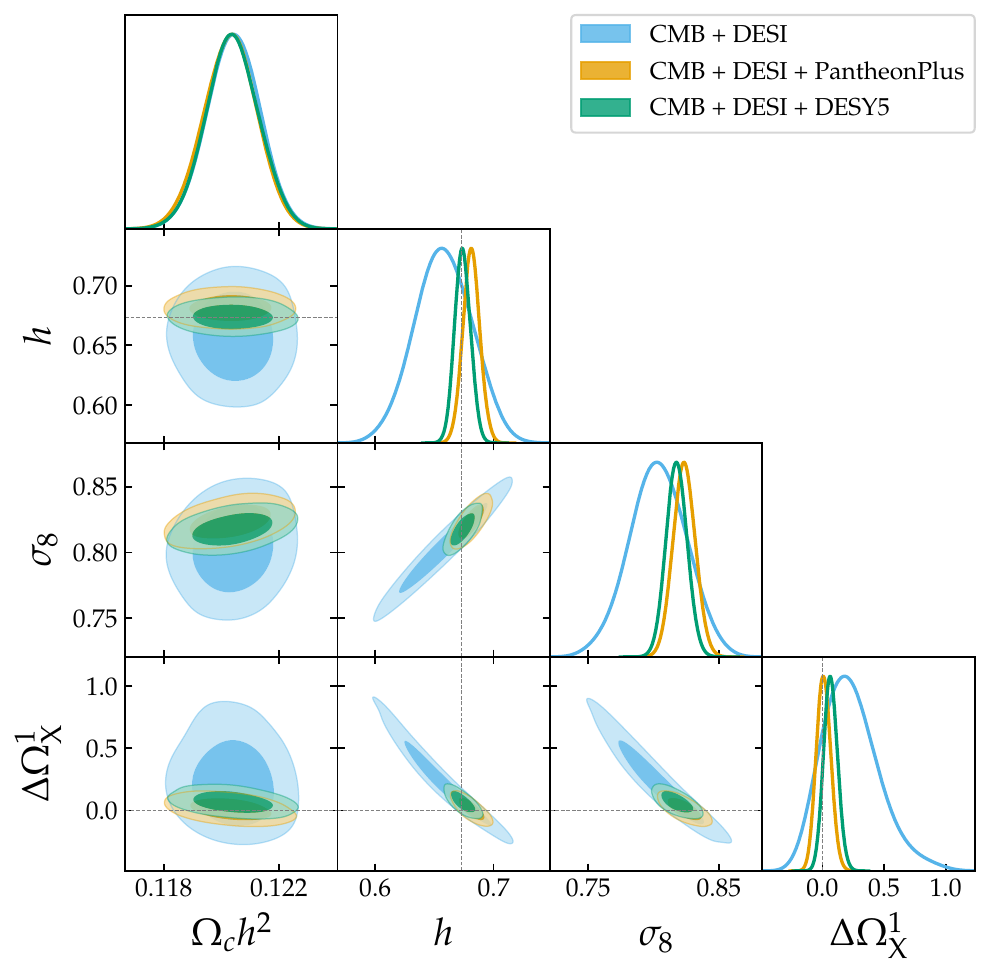}
\caption{\label{fig:contour_3} Results for the 4-nodes case: joint constraints (68\% and 95\% confidence regions) and marginalized posterior distributions on the Hubble parameter and the lowest redshift nodes of $\Delta\Omega_{\rm X}(z)$ (Eq.~\ref{eq:omega_def}), corresponding to the redshifts $z_{1,2,3} = \{0.8,\ 1.6,\ 2.4\}$. Gray dotted lines mark the $\Lambda$CDM limit. 
}
\end{figure}
\begin{figure}
\centering
\includegraphics[width=1\columnwidth]{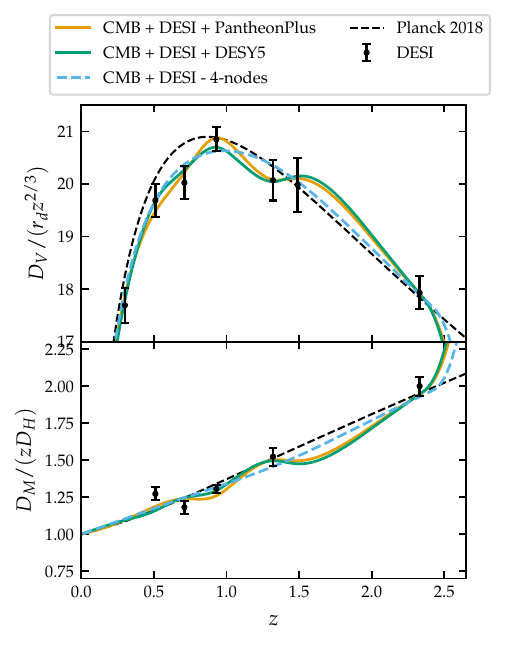}
\caption{\label{fig:distances} Observed distances from BAO measurements, as defined in \autoref{sec:analysis_obs}. We compare the measured data points from DESI with the theoretical prediction for a Planck 2018 cosmology (dashed black line) and with results from the reconstruction. For the 8-nodes case (solid lines), we show the theoretical prediction from the bestfit analysis for CMB + DESI combined with PantheonPlus and DESY5. For the 4-nodes case (dashed blue line), we show bestfit results for CMB + DESI. }
\end{figure}
To assess the robustness of the results from the 8-nodes case presented above and estimate the impact of potential overfitting, we explore an alternative reconstruction with a reduced number of nodes. This test allows us to evaluate the impact of the reconstruction settings on the inferred background evolution. We find that reducing the number of nodes enables us to constrain $\Delta\Omega_{\rm X}(z)$ effectively even using CMB observations combined with DESI alone, which previously led to an unconstrained evolution. Thus, in the following, we discuss results for the 4-nodes case using CMB observations in combination with BAO alone, and BAO with SNe. Results are presented in \autoref{fig:shift_ev_3}, \autoref{fig:contour_3}, \autoref{fig:distances}, and \autoref{tab:constraints_3}.

As expected, for all three combinations of data sets considered, the reconstructed background exhibits a smoother, less wiggly behavior. In the CMB + DESI case, we obtain a significantly low value $h$, although with larger errors with respect to when SNe are included. Although in the second redshift bin we observe a deviation from $\Delta\Omega_{\rm X}(z)=0$ at more than $\sim 2\sigma$ (\autoref{fig:shift_ev_3}, first column), the model is overall compatible with $\Lambda$CDM at the 1.34$\sigma$ level. Indeed, in \autoref{fig:contour_3} we notice that CMB + DESI data alone push the constraint on $\Delta\Omega_{\rm X}^2$ away from its $\Lambda$CDM limit.
Moreover, we observe that $h$ and $\Delta\Omega_{\rm X}^1$ appear to be strongly correlated. 

Including SNe data leads to significantly tighter constraints, in particular on the first bin $\Delta\Omega_{\rm X}^1$ and the cosmological parameters. However, from \autoref{fig:shift_ev_3}, we observe that all three combinations of data sets result in a qualitatively very similar evolution of both $\Delta\Omega_{\rm X}(z)$ and $w(z)$. As for the 8-nodes case, the latter exhibits a phantom-crossing around $z\sim0.6$, but without any wiggles at lower redshifts. In this case, the form of $w(z)$ is much less similar to results found in \cite{Ye:2024ywg}, but it resembles the $w(z)$ obtained with other reconstruction techniques, e.g.\ see Figure 1 in \cite{DESI:2024crossing}.

Again, we find that the result most in tension with $\Lambda$CDM is obtained using DESY5 observations. The computed significance of the deviation is $2.66\sigma$ for DESY5, against the $0.98\sigma$ found for PantheonPlus, which is interestingly less in tension than the CMB + DESI only case. In agreement with the results of the previous section, the tension arising with DESY5 data seems to be a consequence of the constraints on the lowest redshift node. In the third column of \autoref{fig:shift_ev_3}, we notice how DESY5 prefers a $\Delta\Omega_{\rm X}(z)$ away from 0 at more than 2$\sigma$ for $z<0.6$. Notably, the constraints in the second bin at $z=1.6$ are, instead, very similar across all cases. This is reasonable considering the redshift range spanned by SNe observations (see \autoref{fig:mus}). However, our results seem to suggest that part of the tension is driven by DESI data alone, and it is not just a result of the combination between DESI and SNe. {Finally, let us note that contrary to the case with 8 nodes, $\Delta\Omega_{\rm X}$ is never smaller than -1 in this case.}

To analyze more deeply the differences between the 8-nodes and 4-nodes case, in \autoref{fig:distances} we compare the measured DESI BAO distances with the theoretical predictions for a Planck 2018 cosmology and the bestfit evolution from the reconstructed density. We show results for the combination of data sets CMB + DESI + PantheonPlus and CMB + DESI + DESY5, for the 8-nodes case, and CMB + DESI, for the 4-nodes one. We notice that for both cases, the results from the reconstructed background are generally more in agreement with the DESI measured data points with respect to Planck 2018. We observe, however, that in the 4-nodes case the redshift evolution of the distances is qualitatively much more similar to the Planck 2018 one. For the 8-nodes, instead, we recover a very peculiar wiggly behavior, a clear consequence of the freedom allowed by the spline reconstruction and a possible indication of overfitting. The requirement for a more standard behavior of the observables could thus provide yet another indicator to infer the possible overfitting. 

In conclusion, we find that the results for the 8-node case are qualitatively in agreement with the 4-node case. However, the significance of the variation from $\Lambda$CDM appears to be sensitive to the number of nodes used in the background reconstruction. With a higher number of nodes, we are able to observe a more detailed differentiation across cases, whereas fewer bins lead to smoother functions and a more uniform evolution. It is, thus, non-trivial to determine the optimal reconstruction settings given the trade-off between the richness of observable features and the robustness of the constraints. We leave the implementation of well-known smoothing techniques (see e.g.\ \cite{Crittenden:2011aa, Pogosian:2021mcs}) to further investigate the robustness of our results for future work.

\section{Conclusions}\label{sec:conclusions}
The BAO measurements for the first year of observations of DESI provided 
{tantalizing suggestion} for dynamical dark energy within the CPL parametrization from combined CMB, BAO, and SNe data \cite{DESI:2024cosmo_bao}. In the past months, the scientific community has put an extensive effort into investigating further the results found by the DESI collaboration, in the quest to constrain the expansion history of the Universe. Among other approaches, one has been to exploit model-independent reconstruction frameworks to test the DE equation of state limiting theory-induced biases.

In this work, 
{we implemented} a new pipeline for non-parametric, model-independent reconstruction of the DE
density and 
{to test} different scenarios. We target the deviations from a cosmological constant contribution of the density parameter $\Delta\Omega_{\rm X}(z)$, for a general exotic DE fluid $\rm X$. Reconstructing the DE density directly instead of the equation of state parameter provides a more straightforward connection with the distances measured by DESI BAO. By means of an MCMC analysis, we constrain from data the value of $\Delta\Omega_{\rm X}(z)$ at a fixed number of nodes. In order to assess the risk of potential overfitting, we consider two different configurations: one with 8 nodes at the redshifts corresponding to the DESI observations (\textit{8-nodes} case), one with 4 nodes equispaced in the range $z=0-2.4$ (\textit{4-nodes} case). Using a third-degree spline interpolation among the nodes, we recover the full evolution of $\Delta\Omega_{\rm X}(z)$. We choose to anchor the value of the DE density today to be $\Delta\Omega_{\rm X}(0) = 0$, while we impose $\Delta\Omega_{\rm X}(z)$ to be constant after $z=4$.

Following \cite{DESI:2024cosmo_bao} and succeeding works, we constrain $\Delta\Omega_{\rm X}(z)$ using CMB observations from Planck and ACT combined with DESI BAO measurements together with PantheonPlus and DESY5 SNe catalogs. Inspired by the concerns raised in \cite{Efstathiou:2024xcq}, we additionally test the constraining power on the background evolution of the low redshift ($z<0.1$) samples in the DESY5 data set, to investigate the possibility of unknown systematic effects. 

The results from our analysis can be summarized as follows:
\begin{itemize}
    \item In this work, we develop a new framework to reconstruct the background DE evolution. With the analysis presented above, we validate our pipeline finding results compatible with the ones obtained from other widely tested codes, such as~\cite{Ye:2024ywg}.
    \item Our results are qualitatively in agreement with the DESI BAO analysis~\cite{DESI:2024cosmo_bao}. Although milder, we find yet further evidence for a non-standard DE background behaviour, with the significance of the deviation being the highest when using DESY5 observations combined with CMB and DESI data ($2.42\sigma$).
    \item In agreement with the results from other analyses (see e.g. \cite{DESI:2024crossing, Ye:2024ywg}), we find that the derived reconstructed $w(z)$ crosses the phantom divide in all the cases we consider. 
    \item With respect to the CPL parametrization, our reconstruction framework allows us to appreciate more redshift-dependent features. We find that the tension with $\Lambda$CDM observed when using DESY5 ($2.42\sigma$) is driven only by the constraint on $\Delta\Omega_{\rm X}(z)$ in the lowest node at $z=0.3$. While overall PantheonPlus prefers a background evolution more in agreement with $\Lambda$CDM ($1.33\sigma$) 
    , the constraints on a few nodes at higher redshifts are less compatible with $\Delta\Omega_{\rm X} = 0$ using PantheonPlus than DESY5, i.e.\ at redshifts $z=0.51,\, 0.93$.
    \item Following \cite{Efstathiou:2024xcq}, when we shift the low redshift ($z<0.1$) samples in the DESY5 data set, we find that the reconstructed background is perfectly in agreement with a cosmological constant contribution and the tension is completely removed. However, we observe that the $\Lambda$CDM limit is recovered at the expense of the constraints on the cosmological parameter being less compatible with a Planck 2018 cosmology, in particular in the value of $h$. Although we acknowledge that no issue was found with the DESY5 data set~\cite{DES:2025tir}, our analysis seems to support the results from \cite{Efstathiou:2024xcq} as a way to reconcile the data with a cosmological constant. 
    \item Varying the number of nodes used for the interpolation qualitatively leads to similar results, although differences in the reconstructed background can be appreciated. In the 4-nodes case, we recover a less oscillating behaviour of both $\Delta\Omega_{\rm X}$ and $w(z)$ and a similar evolution for all the combinations of data sets considered. As for the 8-nodes case, the most significant deviation from $\Lambda$CDM is obtained with DESY5 data ($2.66\sigma$). We find that $w(z)$ crosses the phantom divide also using this configuration. 
    \item In the 4-nodes case, the reduced number of parameters allows to achieve convergence also with CMB observations combined only with DESI BAO measurements. We can observe how, even without the addition of the SNe data sets, the DESI data push the evolution away from the $\Lambda$CDM limit, e.g. in the second node at $z=1.6$. Adding the SNe reduces significantly the error on the first node at $z=0.8$ and on $h$, with PantheonPlus and DESY5 reducing and increasing the tension, respectively.
\end{itemize}

The pipeline developed, tested, and used here is ready to be applied to future high-precision data. The analysis we present here offers several prospects for ensuing projects. Indeed, we leave for future work the implementation of more sophisticated overfitting control techniques and the extension to the reconstruction of more complex DE and MG models. 
The combination of improved data with flexible analysis will ultimately elucidate whether dark energy is dynamical, as well as potential implications for such a discovery.

\section*{Data Availability}
Access to the original code is available upon reasonable request to the corresponding author.

\begin{acknowledgments}
The authors would like to thank J. Carron, S.B. Haridasu, S. Castello, S. Nesseris, and A. Chudaykin for useful discussion and feedback.

MB, CB and MK acknowledge support from the Swiss National Science Foundation. CB is also supported by the European Research Council (ERC) under the European Union's Horizon 2020 research and innovation program (grant agreement No.~863929; project title ``Testing the law of gravity with novel large-scale structure observables'' EB is supported by the European Union’s Horizon Europe research and innovation programme under the Marie Sklodowska-Curie Postdoctoral Fellowship Programme, SMASH co-funded under the grant agreement No. 101081355. MV is partly supported by the INDARK INFN grant and by the Fondazione ICSC, Spoke 3 ``Astrophysics and Cosmos Observations'', Piano Nazionale di Ripresa e Resilienza Project ID CN00000013 ``Italian Research Center on High-Performance Computing, Big Data and Quantum Computing'' funded by MUR Missione 4 Componente 2 Investimento 1.4: Potenziamento strutture di ricerca e creazione di ``campioni nazionali di R\&S (M4C2-19)'' - Next Generation EU (NGEU).

\end{acknowledgments}

\bibliography{apssamp}

\begin{thebibliography}{71}%
\makeatletter
\providecommand \@ifxundefined [1]{%
 \@ifx{#1\undefined}
}%
\providecommand \@ifnum [1]{%
 \ifnum #1\expandafter \@firstoftwo
 \else \expandafter \@secondoftwo
 \fi
}%
\providecommand \@ifx [1]{%
 \ifx #1\expandafter \@firstoftwo
 \else \expandafter \@secondoftwo
 \fi
}%
\providecommand \natexlab [1]{#1}%
\providecommand \enquote  [1]{``#1''}%
\providecommand \bibnamefont  [1]{#1}%
\providecommand \bibfnamefont [1]{#1}%
\providecommand \citenamefont [1]{#1}%
\providecommand \href@noop [0]{\@secondoftwo}%
\providecommand \href [0]{\begingroup \@sanitize@url \@href}%
\providecommand \@href[1]{\@@startlink{#1}\@@href}%
\providecommand \@@href[1]{\endgroup#1\@@endlink}%
\providecommand \@sanitize@url [0]{\catcode `\\12\catcode `\$12\catcode `\&12\catcode `\#12\catcode `\^12\catcode `\_12\catcode `\%12\relax}%
\providecommand \@@startlink[1]{}%
\providecommand \@@endlink[0]{}%
\providecommand \url  [0]{\begingroup\@sanitize@url \@url }%
\providecommand \@url [1]{\endgroup\@href {#1}{\urlprefix }}%
\providecommand \urlprefix  [0]{URL }%
\providecommand \Eprint [0]{\href }%
\providecommand \doibase [0]{https://doi.org/}%
\providecommand \selectlanguage [0]{\@gobble}%
\providecommand \bibinfo  [0]{\@secondoftwo}%
\providecommand \bibfield  [0]{\@secondoftwo}%
\providecommand \translation [1]{[#1]}%
\providecommand \BibitemOpen [0]{}%
\providecommand \bibitemStop [0]{}%
\providecommand \bibitemNoStop [0]{.\EOS\space}%
\providecommand \EOS [0]{\spacefactor3000\relax}%
\providecommand \BibitemShut  [1]{\csname bibitem#1\endcsname}%
\let\auto@bib@innerbib\@empty
\bibitem [{\citenamefont {Riess}\ \emph {et~al.}(1998)\citenamefont {Riess} \emph {et~al.}}]{SupernovaSearchTeam:1998fmf}%
  \BibitemOpen
  \bibfield  {author} {\bibinfo {author} {\bibfnamefont {A.~G.}\ \bibnamefont {Riess}} \emph {et~al.} (\bibinfo {collaboration} {Supernova Search Team}),\ }\bibfield  {title} {\bibinfo {title} {{Observational evidence from supernovae for an accelerating universe and a cosmological constant}},\ }\href {https://doi.org/10.1086/300499} {\bibfield  {journal} {\bibinfo  {journal} {Astron. J.}\ }\textbf {\bibinfo {volume} {116}},\ \bibinfo {pages} {1009} (\bibinfo {year} {1998})},\ \Eprint {https://arxiv.org/abs/astro-ph/9805201} {arXiv:astro-ph/9805201} \BibitemShut {NoStop}%
\bibitem [{\citenamefont {Perlmutter}\ \emph {et~al.}(1999)\citenamefont {Perlmutter} \emph {et~al.}}]{SupernovaCosmologyProject:1998vns}%
  \BibitemOpen
  \bibfield  {author} {\bibinfo {author} {\bibfnamefont {S.}~\bibnamefont {Perlmutter}} \emph {et~al.} (\bibinfo {collaboration} {Supernova Cosmology Project}),\ }\bibfield  {title} {\bibinfo {title} {{Measurements of $\Omega$ and $\Lambda$ from 42 High Redshift Supernovae}},\ }\href {https://doi.org/10.1086/307221} {\bibfield  {journal} {\bibinfo  {journal} {Astrophys. J.}\ }\textbf {\bibinfo {volume} {517}},\ \bibinfo {pages} {565} (\bibinfo {year} {1999})},\ \Eprint {https://arxiv.org/abs/astro-ph/9812133} {arXiv:astro-ph/9812133} \BibitemShut {NoStop}%
\bibitem [{\citenamefont {Ishak}(2019)}]{Ishak:2018his}%
  \BibitemOpen
  \bibfield  {author} {\bibinfo {author} {\bibfnamefont {M.}~\bibnamefont {Ishak}},\ }\bibfield  {title} {\bibinfo {title} {{Testing General Relativity in Cosmology}},\ }\href {https://doi.org/10.1007/s41114-018-0017-4} {\bibfield  {journal} {\bibinfo  {journal} {Living Rev. Rel.}\ }\textbf {\bibinfo {volume} {22}},\ \bibinfo {pages} {1} (\bibinfo {year} {2019})},\ \Eprint {https://arxiv.org/abs/1806.10122} {arXiv:1806.10122 [astro-ph.CO]} \BibitemShut {NoStop}%
\bibitem [{\citenamefont {Ferreira}(2019)}]{Ferreira:2019xrr}%
  \BibitemOpen
  \bibfield  {author} {\bibinfo {author} {\bibfnamefont {P.~G.}\ \bibnamefont {Ferreira}},\ }\bibfield  {title} {\bibinfo {title} {{Cosmological Tests of Gravity}},\ }\href {https://doi.org/10.1146/annurev-astro-091918-104423} {\bibfield  {journal} {\bibinfo  {journal} {Ann. Rev. Astron. Astrophys.}\ }\textbf {\bibinfo {volume} {57}},\ \bibinfo {pages} {335} (\bibinfo {year} {2019})},\ \Eprint {https://arxiv.org/abs/1902.10503} {arXiv:1902.10503 [astro-ph.CO]} \BibitemShut {NoStop}%
\bibitem [{\citenamefont {Hou}\ \emph {et~al.}(2023)\citenamefont {Hou}, \citenamefont {Bautista}, \citenamefont {Berti}, \citenamefont {Cuesta-Lazaro}, \citenamefont {Hern\'andez-Aguayo}, \citenamefont {Tr\"oster},\ and\ \citenamefont {Zheng}}]{Hou:2023kfp}%
  \BibitemOpen
  \bibfield  {author} {\bibinfo {author} {\bibfnamefont {J.}~\bibnamefont {Hou}}, \bibinfo {author} {\bibfnamefont {J.}~\bibnamefont {Bautista}}, \bibinfo {author} {\bibfnamefont {M.}~\bibnamefont {Berti}}, \bibinfo {author} {\bibfnamefont {C.}~\bibnamefont {Cuesta-Lazaro}}, \bibinfo {author} {\bibfnamefont {C.}~\bibnamefont {Hern\'andez-Aguayo}}, \bibinfo {author} {\bibfnamefont {T.}~\bibnamefont {Tr\"oster}},\ and\ \bibinfo {author} {\bibfnamefont {J.}~\bibnamefont {Zheng}},\ }\bibfield  {title} {\bibinfo {title} {{Cosmological Probes of Structure Growth and Tests of Gravity}},\ }\href {https://doi.org/10.3390/universe9070302} {\bibfield  {journal} {\bibinfo  {journal} {Universe}\ }\textbf {\bibinfo {volume} {9}},\ \bibinfo {pages} {302} (\bibinfo {year} {2023})},\ \Eprint {https://arxiv.org/abs/2306.13726} {arXiv:2306.13726 [gr-qc]} \BibitemShut {NoStop}%
\bibitem [{\citenamefont {Bertschinger}(2006)}]{Bertschinger:2006aw}%
  \BibitemOpen
  \bibfield  {author} {\bibinfo {author} {\bibfnamefont {E.}~\bibnamefont {Bertschinger}},\ }\bibfield  {title} {\bibinfo {title} {{On the Growth of Perturbations as a Test of Dark Energy}},\ }\href {https://doi.org/10.1086/506021} {\bibfield  {journal} {\bibinfo  {journal} {Astrophys. J.}\ }\textbf {\bibinfo {volume} {648}},\ \bibinfo {pages} {797} (\bibinfo {year} {2006})},\ \Eprint {https://arxiv.org/abs/astro-ph/0604485} {arXiv:astro-ph/0604485} \BibitemShut {NoStop}%
\bibitem [{\citenamefont {Amendola}\ \emph {et~al.}(2008)\citenamefont {Amendola}, \citenamefont {Kunz},\ and\ \citenamefont {Sapone}}]{Amendola:2007}%
  \BibitemOpen
  \bibfield  {author} {\bibinfo {author} {\bibfnamefont {L.}~\bibnamefont {Amendola}}, \bibinfo {author} {\bibfnamefont {M.}~\bibnamefont {Kunz}},\ and\ \bibinfo {author} {\bibfnamefont {D.}~\bibnamefont {Sapone}},\ }\bibfield  {title} {\bibinfo {title} {{Measuring the dark side (with weak lensing)}},\ }\href {https://doi.org/10.1088/1475-7516/2008/04/013} {\bibfield  {journal} {\bibinfo  {journal} {JCAP}\ }\textbf {\bibinfo {volume} {04}},\ \bibinfo {pages} {013}},\ \Eprint {https://arxiv.org/abs/0704.2421} {arXiv:0704.2421 [astro-ph]} \BibitemShut {NoStop}%
\bibitem [{\citenamefont {Pogosian}\ and\ \citenamefont {Silvestri}(2008)}]{Pogosian:2007sw}%
  \BibitemOpen
  \bibfield  {author} {\bibinfo {author} {\bibfnamefont {L.}~\bibnamefont {Pogosian}}\ and\ \bibinfo {author} {\bibfnamefont {A.}~\bibnamefont {Silvestri}},\ }\bibfield  {title} {\bibinfo {title} {{The pattern of growth in viable f(R) cosmologies}},\ }\href {https://doi.org/10.1103/PhysRevD.77.023503} {\bibfield  {journal} {\bibinfo  {journal} {Phys. Rev. D}\ }\textbf {\bibinfo {volume} {77}},\ \bibinfo {pages} {023503} (\bibinfo {year} {2008})},\ \bibinfo {note} {[Erratum: Phys.Rev.D 81, 049901 (2010)]},\ \Eprint {https://arxiv.org/abs/0709.0296} {arXiv:0709.0296 [astro-ph]} \BibitemShut {NoStop}%
\bibitem [{\citenamefont {Zhao}\ \emph {et~al.}(2009{\natexlab{a}})\citenamefont {Zhao}, \citenamefont {Pogosian}, \citenamefont {Silvestri},\ and\ \citenamefont {Zylberberg}}]{Zhao:2008}%
  \BibitemOpen
  \bibfield  {author} {\bibinfo {author} {\bibfnamefont {G.-B.}\ \bibnamefont {Zhao}}, \bibinfo {author} {\bibfnamefont {L.}~\bibnamefont {Pogosian}}, \bibinfo {author} {\bibfnamefont {A.}~\bibnamefont {Silvestri}},\ and\ \bibinfo {author} {\bibfnamefont {J.}~\bibnamefont {Zylberberg}},\ }\bibfield  {title} {\bibinfo {title} {{Searching for modified growth patterns with tomographic surveys}},\ }\href {https://doi.org/10.1103/PhysRevD.79.083513} {\bibfield  {journal} {\bibinfo  {journal} {Phys. Rev. D}\ }\textbf {\bibinfo {volume} {79}},\ \bibinfo {pages} {083513} (\bibinfo {year} {2009}{\natexlab{a}})},\ \Eprint {https://arxiv.org/abs/0809.3791} {arXiv:0809.3791 [astro-ph]} \BibitemShut {NoStop}%
\bibitem [{\citenamefont {Zhang}\ \emph {et~al.}(2007)\citenamefont {Zhang}, \citenamefont {Liguori}, \citenamefont {Bean},\ and\ \citenamefont {Dodelson}}]{Zhang:2007}%
  \BibitemOpen
  \bibfield  {author} {\bibinfo {author} {\bibfnamefont {P.}~\bibnamefont {Zhang}}, \bibinfo {author} {\bibfnamefont {M.}~\bibnamefont {Liguori}}, \bibinfo {author} {\bibfnamefont {R.}~\bibnamefont {Bean}},\ and\ \bibinfo {author} {\bibfnamefont {S.}~\bibnamefont {Dodelson}},\ }\bibfield  {title} {\bibinfo {title} {{Probing Gravity at Cosmological Scales by Measurements which Test the Relationship between Gravitational Lensing and Matter Overdensity}},\ }\href {https://doi.org/10.1103/PhysRevLett.99.141302} {\bibfield  {journal} {\bibinfo  {journal} {Phys. Rev. Lett.}\ }\textbf {\bibinfo {volume} {99}},\ \bibinfo {pages} {141302} (\bibinfo {year} {2007})},\ \Eprint {https://arxiv.org/abs/0704.1932} {arXiv:0704.1932 [astro-ph]} \BibitemShut {NoStop}%
\bibitem [{\citenamefont {Bloomfield}\ \emph {et~al.}(2013)\citenamefont {Bloomfield}, \citenamefont {Flanagan}, \citenamefont {Park},\ and\ \citenamefont {Watson}}]{Bloomfield:2012ff}%
  \BibitemOpen
  \bibfield  {author} {\bibinfo {author} {\bibfnamefont {J.~K.}\ \bibnamefont {Bloomfield}}, \bibinfo {author} {\bibfnamefont {E.~E.}\ \bibnamefont {Flanagan}}, \bibinfo {author} {\bibfnamefont {M.}~\bibnamefont {Park}},\ and\ \bibinfo {author} {\bibfnamefont {S.}~\bibnamefont {Watson}},\ }\bibfield  {title} {\bibinfo {title} {{Dark energy or modified gravity? An effective field theory approach}},\ }\href {https://doi.org/10.1088/1475-7516/2013/08/010} {\bibfield  {journal} {\bibinfo  {journal} {JCAP}\ }\textbf {\bibinfo {volume} {08}},\ \bibinfo {pages} {010}},\ \Eprint {https://arxiv.org/abs/1211.7054} {arXiv:1211.7054 [astro-ph.CO]} \BibitemShut {NoStop}%
\bibitem [{\citenamefont {Gubitosi}\ \emph {et~al.}(2013)\citenamefont {Gubitosi}, \citenamefont {Piazza},\ and\ \citenamefont {Vernizzi}}]{Gubitosi:2012hu}%
  \BibitemOpen
  \bibfield  {author} {\bibinfo {author} {\bibfnamefont {G.}~\bibnamefont {Gubitosi}}, \bibinfo {author} {\bibfnamefont {F.}~\bibnamefont {Piazza}},\ and\ \bibinfo {author} {\bibfnamefont {F.}~\bibnamefont {Vernizzi}},\ }\bibfield  {title} {\bibinfo {title} {{The Effective Field Theory of Dark Energy}},\ }\href {https://doi.org/10.1088/1475-7516/2013/02/032} {\bibfield  {journal} {\bibinfo  {journal} {JCAP}\ }\textbf {\bibinfo {volume} {02}},\ \bibinfo {pages} {032}},\ \Eprint {https://arxiv.org/abs/1210.0201} {arXiv:1210.0201 [hep-th]} \BibitemShut {NoStop}%
\bibitem [{\citenamefont {Bellini}\ and\ \citenamefont {Sawicki}(2014)}]{Bellini:2014fua}%
  \BibitemOpen
  \bibfield  {author} {\bibinfo {author} {\bibfnamefont {E.}~\bibnamefont {Bellini}}\ and\ \bibinfo {author} {\bibfnamefont {I.}~\bibnamefont {Sawicki}},\ }\bibfield  {title} {\bibinfo {title} {{Maximal freedom at minimum cost: linear large-scale structure in general modifications of gravity}},\ }\href {https://doi.org/10.1088/1475-7516/2014/07/050} {\bibfield  {journal} {\bibinfo  {journal} {JCAP}\ }\textbf {\bibinfo {volume} {07}},\ \bibinfo {pages} {050}},\ \Eprint {https://arxiv.org/abs/1404.3713} {arXiv:1404.3713 [astro-ph.CO]} \BibitemShut {NoStop}%
\bibitem [{\citenamefont {Huterer}\ and\ \citenamefont {Starkman}(2003)}]{Huterer:2002hy}%
  \BibitemOpen
  \bibfield  {author} {\bibinfo {author} {\bibfnamefont {D.}~\bibnamefont {Huterer}}\ and\ \bibinfo {author} {\bibfnamefont {G.}~\bibnamefont {Starkman}},\ }\bibfield  {title} {\bibinfo {title} {{Parameterization of dark-energy properties: A Principal-component approach}},\ }\href {https://doi.org/10.1103/PhysRevLett.90.031301} {\bibfield  {journal} {\bibinfo  {journal} {Phys. Rev. Lett.}\ }\textbf {\bibinfo {volume} {90}},\ \bibinfo {pages} {031301} (\bibinfo {year} {2003})},\ \Eprint {https://arxiv.org/abs/astro-ph/0207517} {arXiv:astro-ph/0207517} \BibitemShut {NoStop}%
\bibitem [{\citenamefont {Mortonson}\ \emph {et~al.}(2009)\citenamefont {Mortonson}, \citenamefont {Hu},\ and\ \citenamefont {Huterer}}]{Mortonson:2008qy}%
  \BibitemOpen
  \bibfield  {author} {\bibinfo {author} {\bibfnamefont {M.~J.}\ \bibnamefont {Mortonson}}, \bibinfo {author} {\bibfnamefont {W.}~\bibnamefont {Hu}},\ and\ \bibinfo {author} {\bibfnamefont {D.}~\bibnamefont {Huterer}},\ }\bibfield  {title} {\bibinfo {title} {{Falsifying Paradigms for Cosmic Acceleration}},\ }\href {https://doi.org/10.1103/PhysRevD.79.023004} {\bibfield  {journal} {\bibinfo  {journal} {Phys. Rev. D}\ }\textbf {\bibinfo {volume} {79}},\ \bibinfo {pages} {023004} (\bibinfo {year} {2009})},\ \Eprint {https://arxiv.org/abs/0810.1744} {arXiv:0810.1744 [astro-ph]} \BibitemShut {NoStop}%
\bibitem [{\citenamefont {Zhao}\ \emph {et~al.}(2009{\natexlab{b}})\citenamefont {Zhao}, \citenamefont {Pogosian}, \citenamefont {Silvestri},\ and\ \citenamefont {Zylberberg}}]{Zhao:2009fn}%
  \BibitemOpen
  \bibfield  {author} {\bibinfo {author} {\bibfnamefont {G.-B.}\ \bibnamefont {Zhao}}, \bibinfo {author} {\bibfnamefont {L.}~\bibnamefont {Pogosian}}, \bibinfo {author} {\bibfnamefont {A.}~\bibnamefont {Silvestri}},\ and\ \bibinfo {author} {\bibfnamefont {J.}~\bibnamefont {Zylberberg}},\ }\bibfield  {title} {\bibinfo {title} {{Cosmological Tests of General Relativity with Future Tomographic Surveys}},\ }\href {https://doi.org/10.1103/PhysRevLett.103.241301} {\bibfield  {journal} {\bibinfo  {journal} {Phys. Rev. Lett.}\ }\textbf {\bibinfo {volume} {103}},\ \bibinfo {pages} {241301} (\bibinfo {year} {2009}{\natexlab{b}})},\ \Eprint {https://arxiv.org/abs/0905.1326} {arXiv:0905.1326 [astro-ph.CO]} \BibitemShut {NoStop}%
\bibitem [{\citenamefont {Zhao}\ \emph {et~al.}(2010)\citenamefont {Zhao}, \citenamefont {Giannantonio}, \citenamefont {Pogosian}, \citenamefont {Silvestri}, \citenamefont {Bacon}, \citenamefont {Koyama}, \citenamefont {Nichol},\ and\ \citenamefont {Song}}]{Zhao:2010dz}%
  \BibitemOpen
  \bibfield  {author} {\bibinfo {author} {\bibfnamefont {G.-B.}\ \bibnamefont {Zhao}}, \bibinfo {author} {\bibfnamefont {T.}~\bibnamefont {Giannantonio}}, \bibinfo {author} {\bibfnamefont {L.}~\bibnamefont {Pogosian}}, \bibinfo {author} {\bibfnamefont {A.}~\bibnamefont {Silvestri}}, \bibinfo {author} {\bibfnamefont {D.~J.}\ \bibnamefont {Bacon}}, \bibinfo {author} {\bibfnamefont {K.}~\bibnamefont {Koyama}}, \bibinfo {author} {\bibfnamefont {R.~C.}\ \bibnamefont {Nichol}},\ and\ \bibinfo {author} {\bibfnamefont {Y.-S.}\ \bibnamefont {Song}},\ }\bibfield  {title} {\bibinfo {title} {{Probing modifications of General Relativity using current cosmological observations}},\ }\href {https://doi.org/10.1103/PhysRevD.81.103510} {\bibfield  {journal} {\bibinfo  {journal} {Phys. Rev. D}\ }\textbf {\bibinfo {volume} {81}},\ \bibinfo {pages} {103510} (\bibinfo {year} {2010})},\ \Eprint {https://arxiv.org/abs/1003.0001} {arXiv:1003.0001 [astro-ph.CO]} \BibitemShut {NoStop}%
\bibitem [{\citenamefont {Mortonson}\ \emph {et~al.}(2010)\citenamefont {Mortonson}, \citenamefont {Hu},\ and\ \citenamefont {Huterer}}]{Mortonson_2010}%
  \BibitemOpen
  \bibfield  {author} {\bibinfo {author} {\bibfnamefont {M.~J.}\ \bibnamefont {Mortonson}}, \bibinfo {author} {\bibfnamefont {W.}~\bibnamefont {Hu}},\ and\ \bibinfo {author} {\bibfnamefont {D.}~\bibnamefont {Huterer}},\ }\bibfield  {title} {\bibinfo {title} {Testable dark energy predictions from current data},\ }\bibfield  {journal} {\bibinfo  {journal} {Physical Review D}\ }\textbf {\bibinfo {volume} {81}},\ \href {https://doi.org/10.1103/physrevd.81.063007} {10.1103/physrevd.81.063007} (\bibinfo {year} {2010})\BibitemShut {NoStop}%
\bibitem [{\citenamefont {Hojjati}\ \emph {et~al.}(2012)\citenamefont {Hojjati}, \citenamefont {Zhao}, \citenamefont {Pogosian}, \citenamefont {Silvestri}, \citenamefont {Crittenden},\ and\ \citenamefont {Koyama}}]{Hojjati:2011xd}%
  \BibitemOpen
  \bibfield  {author} {\bibinfo {author} {\bibfnamefont {A.}~\bibnamefont {Hojjati}}, \bibinfo {author} {\bibfnamefont {G.-B.}\ \bibnamefont {Zhao}}, \bibinfo {author} {\bibfnamefont {L.}~\bibnamefont {Pogosian}}, \bibinfo {author} {\bibfnamefont {A.}~\bibnamefont {Silvestri}}, \bibinfo {author} {\bibfnamefont {R.}~\bibnamefont {Crittenden}},\ and\ \bibinfo {author} {\bibfnamefont {K.}~\bibnamefont {Koyama}},\ }\bibfield  {title} {\bibinfo {title} {{Cosmological tests of General Relativity: a principal component analysis}},\ }\href {https://doi.org/10.1103/PhysRevD.85.043508} {\bibfield  {journal} {\bibinfo  {journal} {Phys. Rev. D}\ }\textbf {\bibinfo {volume} {85}},\ \bibinfo {pages} {043508} (\bibinfo {year} {2012})},\ \Eprint {https://arxiv.org/abs/1111.3960} {arXiv:1111.3960 [astro-ph.CO]} \BibitemShut {NoStop}%
\bibitem [{\citenamefont {Zhao}\ \emph {et~al.}(2012)\citenamefont {Zhao}, \citenamefont {Crittenden}, \citenamefont {Pogosian},\ and\ \citenamefont {Zhang}}]{Zhao_2012}%
  \BibitemOpen
  \bibfield  {author} {\bibinfo {author} {\bibfnamefont {G.-B.}\ \bibnamefont {Zhao}}, \bibinfo {author} {\bibfnamefont {R.~G.}\ \bibnamefont {Crittenden}}, \bibinfo {author} {\bibfnamefont {L.}~\bibnamefont {Pogosian}},\ and\ \bibinfo {author} {\bibfnamefont {X.}~\bibnamefont {Zhang}},\ }\bibfield  {title} {\bibinfo {title} {Examining the evidence for dynamical dark energy},\ }\bibfield  {journal} {\bibinfo  {journal} {Physical Review Letters}\ }\textbf {\bibinfo {volume} {109}},\ \href {https://doi.org/10.1103/physrevlett.109.171301} {10.1103/physrevlett.109.171301} (\bibinfo {year} {2012})\BibitemShut {NoStop}%
\bibitem [{\citenamefont {Zhao}\ \emph {et~al.}(2017)\citenamefont {Zhao} \emph {et~al.}}]{Zhao:2017cud}%
  \BibitemOpen
  \bibfield  {author} {\bibinfo {author} {\bibfnamefont {G.-B.}\ \bibnamefont {Zhao}} \emph {et~al.},\ }\bibfield  {title} {\bibinfo {title} {{Dynamical dark energy in light of the latest observations}},\ }\href {https://doi.org/10.1038/s41550-017-0216-z} {\bibfield  {journal} {\bibinfo  {journal} {Nature Astron.}\ }\textbf {\bibinfo {volume} {1}},\ \bibinfo {pages} {627} (\bibinfo {year} {2017})},\ \Eprint {https://arxiv.org/abs/1701.08165} {arXiv:1701.08165 [astro-ph.CO]} \BibitemShut {NoStop}%
\bibitem [{\citenamefont {Casas}\ \emph {et~al.}(2017)\citenamefont {Casas}, \citenamefont {Kunz}, \citenamefont {Martinelli},\ and\ \citenamefont {Pettorino}}]{Casas:2017eob}%
  \BibitemOpen
  \bibfield  {author} {\bibinfo {author} {\bibfnamefont {S.}~\bibnamefont {Casas}}, \bibinfo {author} {\bibfnamefont {M.}~\bibnamefont {Kunz}}, \bibinfo {author} {\bibfnamefont {M.}~\bibnamefont {Martinelli}},\ and\ \bibinfo {author} {\bibfnamefont {V.}~\bibnamefont {Pettorino}},\ }\bibfield  {title} {\bibinfo {title} {{Linear and non-linear Modified Gravity forecasts with future surveys}},\ }\href {https://doi.org/10.1016/j.dark.2017.09.009} {\bibfield  {journal} {\bibinfo  {journal} {Phys. Dark Univ.}\ }\textbf {\bibinfo {volume} {18}},\ \bibinfo {pages} {73} (\bibinfo {year} {2017})},\ \Eprint {https://arxiv.org/abs/1703.01271} {arXiv:1703.01271 [astro-ph.CO]} \BibitemShut {NoStop}%
\bibitem [{\citenamefont {Raveri}(2020)}]{Raveri:2019mxg}%
  \BibitemOpen
  \bibfield  {author} {\bibinfo {author} {\bibfnamefont {M.}~\bibnamefont {Raveri}},\ }\bibfield  {title} {\bibinfo {title} {{Reconstructing Gravity on Cosmological Scales}},\ }\href {https://doi.org/10.1103/PhysRevD.101.083524} {\bibfield  {journal} {\bibinfo  {journal} {Phys. Rev. D}\ }\textbf {\bibinfo {volume} {101}},\ \bibinfo {pages} {083524} (\bibinfo {year} {2020})},\ \Eprint {https://arxiv.org/abs/1902.01366} {arXiv:1902.01366 [astro-ph.CO]} \BibitemShut {NoStop}%
\bibitem [{\citenamefont {Park}\ \emph {et~al.}(2021)\citenamefont {Park}, \citenamefont {Raveri},\ and\ \citenamefont {Jain}}]{Park:2021jmi}%
  \BibitemOpen
  \bibfield  {author} {\bibinfo {author} {\bibfnamefont {M.}~\bibnamefont {Park}}, \bibinfo {author} {\bibfnamefont {M.}~\bibnamefont {Raveri}},\ and\ \bibinfo {author} {\bibfnamefont {B.}~\bibnamefont {Jain}},\ }\bibfield  {title} {\bibinfo {title} {{Reconstructing Quintessence}},\ }\href {https://doi.org/10.1103/PhysRevD.103.103530} {\bibfield  {journal} {\bibinfo  {journal} {Phys. Rev. D}\ }\textbf {\bibinfo {volume} {103}},\ \bibinfo {pages} {103530} (\bibinfo {year} {2021})},\ \Eprint {https://arxiv.org/abs/2101.04666} {arXiv:2101.04666 [astro-ph.CO]} \BibitemShut {NoStop}%
\bibitem [{\citenamefont {Pogosian}\ \emph {et~al.}(2022)\citenamefont {Pogosian}, \citenamefont {Raveri}, \citenamefont {Koyama}, \citenamefont {Martinelli}, \citenamefont {Silvestri}, \citenamefont {Zhao}, \citenamefont {Li}, \citenamefont {Peirone},\ and\ \citenamefont {Zucca}}]{Pogosian:2021mcs}%
  \BibitemOpen
  \bibfield  {author} {\bibinfo {author} {\bibfnamefont {L.}~\bibnamefont {Pogosian}}, \bibinfo {author} {\bibfnamefont {M.}~\bibnamefont {Raveri}}, \bibinfo {author} {\bibfnamefont {K.}~\bibnamefont {Koyama}}, \bibinfo {author} {\bibfnamefont {M.}~\bibnamefont {Martinelli}}, \bibinfo {author} {\bibfnamefont {A.}~\bibnamefont {Silvestri}}, \bibinfo {author} {\bibfnamefont {G.-B.}\ \bibnamefont {Zhao}}, \bibinfo {author} {\bibfnamefont {J.}~\bibnamefont {Li}}, \bibinfo {author} {\bibfnamefont {S.}~\bibnamefont {Peirone}},\ and\ \bibinfo {author} {\bibfnamefont {A.}~\bibnamefont {Zucca}},\ }\bibfield  {title} {\bibinfo {title} {{Imprints of cosmological tensions in reconstructed gravity}},\ }\href {https://doi.org/10.1038/s41550-022-01808-7} {\bibfield  {journal} {\bibinfo  {journal} {Nature Astron.}\ }\textbf {\bibinfo {volume} {6}},\ \bibinfo {pages} {1484} (\bibinfo {year} {2022})},\ \Eprint {https://arxiv.org/abs/2107.12992} {arXiv:2107.12992 [astro-ph.CO]} \BibitemShut {NoStop}%
\bibitem [{\citenamefont {Raveri}\ \emph {et~al.}(2023)\citenamefont {Raveri}, \citenamefont {Pogosian}, \citenamefont {Martinelli}, \citenamefont {Koyama}, \citenamefont {Silvestri},\ and\ \citenamefont {Zhao}}]{Raveri:2021dbu}%
  \BibitemOpen
  \bibfield  {author} {\bibinfo {author} {\bibfnamefont {M.}~\bibnamefont {Raveri}}, \bibinfo {author} {\bibfnamefont {L.}~\bibnamefont {Pogosian}}, \bibinfo {author} {\bibfnamefont {M.}~\bibnamefont {Martinelli}}, \bibinfo {author} {\bibfnamefont {K.}~\bibnamefont {Koyama}}, \bibinfo {author} {\bibfnamefont {A.}~\bibnamefont {Silvestri}},\ and\ \bibinfo {author} {\bibfnamefont {G.-B.}\ \bibnamefont {Zhao}},\ }\bibfield  {title} {\bibinfo {title} {{Principal reconstructed modes of dark energy and gravity}},\ }\href {https://doi.org/10.1088/1475-7516/2023/02/061} {\bibfield  {journal} {\bibinfo  {journal} {JCAP}\ }\textbf {\bibinfo {volume} {02}},\ \bibinfo {pages} {061}},\ \Eprint {https://arxiv.org/abs/2107.12990} {arXiv:2107.12990 [astro-ph.CO]} \BibitemShut {NoStop}%
\bibitem [{\citenamefont {Rubin}\ \emph {et~al.}(2023)\citenamefont {Rubin} \emph {et~al.}}]{Rubin:2023ovl}%
  \BibitemOpen
  \bibfield  {author} {\bibinfo {author} {\bibfnamefont {D.}~\bibnamefont {Rubin}} \emph {et~al.},\ }\href@noop {} {\bibinfo {title} {{Union Through UNITY: Cosmology with 2,000 SNe Using a Unified Bayesian Framework}}} (\bibinfo {year} {2023}),\ \Eprint {https://arxiv.org/abs/2311.12098} {arXiv:2311.12098 [astro-ph.CO]} \BibitemShut {NoStop}%
\bibitem [{\citenamefont {Adame}\ \emph {et~al.}(2024{\natexlab{a}})\citenamefont {Adame} \emph {et~al.}}]{DESI:2024cosmo_bao}%
  \BibitemOpen
  \bibfield  {author} {\bibinfo {author} {\bibfnamefont {A.~G.}\ \bibnamefont {Adame}} \emph {et~al.} (\bibinfo {collaboration} {DESI}),\ }\href@noop {} {\bibinfo {title} {{DESI 2024 VI: Cosmological Constraints from the Measurements of Baryon Acoustic Oscillations}}} (\bibinfo {year} {2024}{\natexlab{a}}),\ \Eprint {https://arxiv.org/abs/2404.03002} {arXiv:2404.03002 [astro-ph.CO]} \BibitemShut {NoStop}%
\bibitem [{\citenamefont {Adame}\ \emph {et~al.}(2024{\natexlab{b}})\citenamefont {Adame} \emph {et~al.}}]{DESI:2024uvr}%
  \BibitemOpen
  \bibfield  {author} {\bibinfo {author} {\bibfnamefont {A.~G.}\ \bibnamefont {Adame}} \emph {et~al.} (\bibinfo {collaboration} {DESI}),\ }\href@noop {} {\bibinfo {title} {{DESI 2024 III: Baryon Acoustic Oscillations from Galaxies and Quasars}}} (\bibinfo {year} {2024}{\natexlab{b}}),\ \Eprint {https://arxiv.org/abs/2404.03000} {arXiv:2404.03000 [astro-ph.CO]} \BibitemShut {NoStop}%
\bibitem [{\citenamefont {Adame}\ \emph {et~al.}(2024{\natexlab{c}})\citenamefont {Adame} \emph {et~al.}}]{DESI:2024full_shape}%
  \BibitemOpen
  \bibfield  {author} {\bibinfo {author} {\bibfnamefont {A.~G.}\ \bibnamefont {Adame}} \emph {et~al.} (\bibinfo {collaboration} {DESI}),\ }\href@noop {} {\bibinfo {title} {{DESI 2024 VII: Cosmological Constraints from the Full-Shape Modeling of Clustering Measurements}}} (\bibinfo {year} {2024}{\natexlab{c}}),\ \Eprint {https://arxiv.org/abs/2411.12022} {arXiv:2411.12022 [astro-ph.CO]} \BibitemShut {NoStop}%
\bibitem [{\citenamefont {Calderon}\ \emph {et~al.}(2024)\citenamefont {Calderon} \emph {et~al.}}]{DESI:2024crossing}%
  \BibitemOpen
  \bibfield  {author} {\bibinfo {author} {\bibfnamefont {R.}~\bibnamefont {Calderon}} \emph {et~al.} (\bibinfo {collaboration} {DESI}),\ }\bibfield  {title} {\bibinfo {title} {{DESI 2024: reconstructing dark energy using crossing statistics with DESI DR1 BAO data}},\ }\href {https://doi.org/10.1088/1475-7516/2024/10/048} {\bibfield  {journal} {\bibinfo  {journal} {JCAP}\ }\textbf {\bibinfo {volume} {10}},\ \bibinfo {pages} {048}},\ \Eprint {https://arxiv.org/abs/2405.04216} {arXiv:2405.04216 [astro-ph.CO]} \BibitemShut {NoStop}%
\bibitem [{\citenamefont {Lodha}\ \emph {et~al.}(2025)\citenamefont {Lodha} \emph {et~al.}}]{DESI:2024kob}%
  \BibitemOpen
  \bibfield  {author} {\bibinfo {author} {\bibfnamefont {K.}~\bibnamefont {Lodha}} \emph {et~al.} (\bibinfo {collaboration} {DESI}),\ }\bibfield  {title} {\bibinfo {title} {{DESI 2024: Constraints on physics-focused aspects of dark energy using DESI DR1 BAO data}},\ }\href {https://doi.org/10.1103/PhysRevD.111.023532} {\bibfield  {journal} {\bibinfo  {journal} {Phys. Rev. D}\ }\textbf {\bibinfo {volume} {111}},\ \bibinfo {pages} {023532} (\bibinfo {year} {2025})},\ \Eprint {https://arxiv.org/abs/2405.13588} {arXiv:2405.13588 [astro-ph.CO]} \BibitemShut {NoStop}%
\bibitem [{\citenamefont {Mukherjee}\ and\ \citenamefont {Sen}(2024)}]{Mukherjee:2024ryz}%
  \BibitemOpen
  \bibfield  {author} {\bibinfo {author} {\bibfnamefont {P.}~\bibnamefont {Mukherjee}}\ and\ \bibinfo {author} {\bibfnamefont {A.~A.}\ \bibnamefont {Sen}},\ }\bibfield  {title} {\bibinfo {title} {{Model-independent cosmological inference post DESI DR1 BAO measurements}},\ }\href {https://doi.org/10.1103/PhysRevD.110.123502} {\bibfield  {journal} {\bibinfo  {journal} {Phys. Rev. D}\ }\textbf {\bibinfo {volume} {110}},\ \bibinfo {pages} {123502} (\bibinfo {year} {2024})},\ \Eprint {https://arxiv.org/abs/2405.19178} {arXiv:2405.19178 [astro-ph.CO]} \BibitemShut {NoStop}%
\bibitem [{\citenamefont {Dinda}\ and\ \citenamefont {Maartens}(2025)}]{Dinda:2024ktd}%
  \BibitemOpen
  \bibfield  {author} {\bibinfo {author} {\bibfnamefont {B.~R.}\ \bibnamefont {Dinda}}\ and\ \bibinfo {author} {\bibfnamefont {R.}~\bibnamefont {Maartens}},\ }\bibfield  {title} {\bibinfo {title} {{Model-agnostic assessment of dark energy after DESI DR1 BAO}},\ }\href {https://doi.org/10.1088/1475-7516/2025/01/120} {\bibfield  {journal} {\bibinfo  {journal} {JCAP}\ }\textbf {\bibinfo {volume} {01}},\ \bibinfo {pages} {120}},\ \Eprint {https://arxiv.org/abs/2407.17252} {arXiv:2407.17252 [astro-ph.CO]} \BibitemShut {NoStop}%
\bibitem [{\citenamefont {Cozzumbo}\ \emph {et~al.}(2024)\citenamefont {Cozzumbo}, \citenamefont {Dupletsa}, \citenamefont {Calder\'on}, \citenamefont {Murgia}, \citenamefont {Oganesyan},\ and\ \citenamefont {Branchesi}}]{Cozzumbo:2024vxw}%
  \BibitemOpen
  \bibfield  {author} {\bibinfo {author} {\bibfnamefont {A.}~\bibnamefont {Cozzumbo}}, \bibinfo {author} {\bibfnamefont {U.}~\bibnamefont {Dupletsa}}, \bibinfo {author} {\bibfnamefont {R.}~\bibnamefont {Calder\'on}}, \bibinfo {author} {\bibfnamefont {R.}~\bibnamefont {Murgia}}, \bibinfo {author} {\bibfnamefont {G.}~\bibnamefont {Oganesyan}},\ and\ \bibinfo {author} {\bibfnamefont {M.}~\bibnamefont {Branchesi}},\ }\href@noop {} {\bibinfo {title} {{Model-independent cosmology with joint observations of gravitational waves and $\gamma$-ray bursts}}} (\bibinfo {year} {2024}),\ \Eprint {https://arxiv.org/abs/2411.02490} {arXiv:2411.02490 [astro-ph.CO]} \BibitemShut {NoStop}%
\bibitem [{\citenamefont {Yang}\ \emph {et~al.}(2025)\citenamefont {Yang}, \citenamefont {Wang}, \citenamefont {Li}, \citenamefont {Yuan}, \citenamefont {Ren}, \citenamefont {Saridakis},\ and\ \citenamefont {Cai}}]{Yang:2025kgc}%
  \BibitemOpen
  \bibfield  {author} {\bibinfo {author} {\bibfnamefont {Y.}~\bibnamefont {Yang}}, \bibinfo {author} {\bibfnamefont {Q.}~\bibnamefont {Wang}}, \bibinfo {author} {\bibfnamefont {C.}~\bibnamefont {Li}}, \bibinfo {author} {\bibfnamefont {P.}~\bibnamefont {Yuan}}, \bibinfo {author} {\bibfnamefont {X.}~\bibnamefont {Ren}}, \bibinfo {author} {\bibfnamefont {E.~N.}\ \bibnamefont {Saridakis}},\ and\ \bibinfo {author} {\bibfnamefont {Y.-F.}\ \bibnamefont {Cai}},\ }\href@noop {} {\bibinfo {title} {{Gaussian-process reconstructions and model building of quintom dark energy from latest cosmological observations}}} (\bibinfo {year} {2025}),\ \Eprint {https://arxiv.org/abs/2501.18336} {arXiv:2501.18336 [astro-ph.CO]} \BibitemShut {NoStop}%
\bibitem [{\citenamefont {Liu}\ \emph {et~al.}(2024)\citenamefont {Liu}, \citenamefont {Wang},\ and\ \citenamefont {Zhao}}]{Liu:2024gfy}%
  \BibitemOpen
  \bibfield  {author} {\bibinfo {author} {\bibfnamefont {G.}~\bibnamefont {Liu}}, \bibinfo {author} {\bibfnamefont {Y.}~\bibnamefont {Wang}},\ and\ \bibinfo {author} {\bibfnamefont {W.}~\bibnamefont {Zhao}},\ }\href@noop {} {\bibinfo {title} {{Impact of LRG1 and LRG2 in DESI 2024 BAO data on dark energy evolution}}} (\bibinfo {year} {2024}),\ \Eprint {https://arxiv.org/abs/2407.04385} {arXiv:2407.04385 [astro-ph.CO]} \BibitemShut {NoStop}%
\bibitem [{\citenamefont {L'Huillier}\ \emph {et~al.}(2024)\citenamefont {L'Huillier}, \citenamefont {Mitra}, \citenamefont {Shafieloo}, \citenamefont {Keeley},\ and\ \citenamefont {Koo}}]{LHuillier:2024rmp}%
  \BibitemOpen
  \bibfield  {author} {\bibinfo {author} {\bibfnamefont {B.}~\bibnamefont {L'Huillier}}, \bibinfo {author} {\bibfnamefont {A.}~\bibnamefont {Mitra}}, \bibinfo {author} {\bibfnamefont {A.}~\bibnamefont {Shafieloo}}, \bibinfo {author} {\bibfnamefont {R.~E.}\ \bibnamefont {Keeley}},\ and\ \bibinfo {author} {\bibfnamefont {H.}~\bibnamefont {Koo}},\ }\href@noop {} {\bibinfo {title} {{Litmus tests of the flat $\Lambda$CDM model and model-independent measurement of $H_0r_\mathrm{d}$ with LSST and DESI}}} (\bibinfo {year} {2024}),\ \Eprint {https://arxiv.org/abs/2407.07847} {arXiv:2407.07847 [astro-ph.CO]} \BibitemShut {NoStop}%
\bibitem [{\citenamefont {Jiang}\ \emph {et~al.}(2024)\citenamefont {Jiang}, \citenamefont {Pedrotti}, \citenamefont {da~Costa},\ and\ \citenamefont {Vagnozzi}}]{Jiang:2024xnu}%
  \BibitemOpen
  \bibfield  {author} {\bibinfo {author} {\bibfnamefont {J.-Q.}\ \bibnamefont {Jiang}}, \bibinfo {author} {\bibfnamefont {D.}~\bibnamefont {Pedrotti}}, \bibinfo {author} {\bibfnamefont {S.~S.}\ \bibnamefont {da~Costa}},\ and\ \bibinfo {author} {\bibfnamefont {S.}~\bibnamefont {Vagnozzi}},\ }\bibfield  {title} {\bibinfo {title} {{Nonparametric late-time expansion history reconstruction and implications for the Hubble tension in light of recent DESI and type Ia supernovae data}},\ }\href {https://doi.org/10.1103/PhysRevD.110.123519} {\bibfield  {journal} {\bibinfo  {journal} {Phys. Rev. D}\ }\textbf {\bibinfo {volume} {110}},\ \bibinfo {pages} {123519} (\bibinfo {year} {2024})},\ \Eprint {https://arxiv.org/abs/2408.02365} {arXiv:2408.02365 [astro-ph.CO]} \BibitemShut {NoStop}%
\bibitem [{\citenamefont {Pang}\ \emph {et~al.}(2024)\citenamefont {Pang}, \citenamefont {Zhang},\ and\ \citenamefont {Huang}}]{Pang:2024qyh}%
  \BibitemOpen
  \bibfield  {author} {\bibinfo {author} {\bibfnamefont {Y.-H.}\ \bibnamefont {Pang}}, \bibinfo {author} {\bibfnamefont {X.}~\bibnamefont {Zhang}},\ and\ \bibinfo {author} {\bibfnamefont {Q.-G.}\ \bibnamefont {Huang}},\ }\href@noop {} {\bibinfo {title} {{Constraints on Redshift-Binned Dark Energy using DESI BAO Data}}} (\bibinfo {year} {2024}),\ \Eprint {https://arxiv.org/abs/2408.14787} {arXiv:2408.14787 [astro-ph.CO]} \BibitemShut {NoStop}%
\bibitem [{\citenamefont {Ye}\ \emph {et~al.}(2024)\citenamefont {Ye}, \citenamefont {Martinelli}, \citenamefont {Hu},\ and\ \citenamefont {Silvestri}}]{Ye:2024ywg}%
  \BibitemOpen
  \bibfield  {author} {\bibinfo {author} {\bibfnamefont {G.}~\bibnamefont {Ye}}, \bibinfo {author} {\bibfnamefont {M.}~\bibnamefont {Martinelli}}, \bibinfo {author} {\bibfnamefont {B.}~\bibnamefont {Hu}},\ and\ \bibinfo {author} {\bibfnamefont {A.}~\bibnamefont {Silvestri}},\ }\href@noop {} {\bibinfo {title} {{Non-minimally coupled gravity as a physically viable fit to DESI 2024 BAO}}} (\bibinfo {year} {2024}),\ \Eprint {https://arxiv.org/abs/2407.15832} {arXiv:2407.15832 [astro-ph.CO]} \BibitemShut {NoStop}%
\bibitem [{\citenamefont {Ye}(2024)}]{Ye:2024alone}%
  \BibitemOpen
  \bibfield  {author} {\bibinfo {author} {\bibfnamefont {G.}~\bibnamefont {Ye}},\ }\href@noop {} {\bibinfo {title} {{Bridge the Cosmological Tensions with Thawing Gravity}}} (\bibinfo {year} {2024}),\ \Eprint {https://arxiv.org/abs/2411.11743} {arXiv:2411.11743 [astro-ph.CO]} \BibitemShut {NoStop}%
\bibitem [{\citenamefont {Aghanim}\ \emph {et~al.}(2020)\citenamefont {Aghanim} \emph {et~al.}}]{planck:2018}%
  \BibitemOpen
  \bibfield  {author} {\bibinfo {author} {\bibfnamefont {N.}~\bibnamefont {Aghanim}} \emph {et~al.} (\bibinfo {collaboration} {Planck}),\ }\bibfield  {title} {\bibinfo {title} {{Planck 2018 results. VI. Cosmological parameters}},\ }\href {https://doi.org/10.1051/0004-6361/201833910} {\bibfield  {journal} {\bibinfo  {journal} {Astron. Astrophys.}\ }\textbf {\bibinfo {volume} {641}},\ \bibinfo {pages} {A6} (\bibinfo {year} {2020})},\ \bibinfo {note} {[Erratum: Astron.Astrophys. 652, C4 (2021)]},\ \Eprint {https://arxiv.org/abs/1807.06209} {arXiv:1807.06209 [astro-ph.CO]} \BibitemShut {NoStop}%
\bibitem [{\citenamefont {Qu}\ \emph {et~al.}(2024)\citenamefont {Qu} \emph {et~al.}}]{ACT:2023dou}%
  \BibitemOpen
  \bibfield  {author} {\bibinfo {author} {\bibfnamefont {F.~J.}\ \bibnamefont {Qu}} \emph {et~al.} (\bibinfo {collaboration} {ACT}),\ }\bibfield  {title} {\bibinfo {title} {{The Atacama Cosmology Telescope: A Measurement of the DR6 CMB Lensing Power Spectrum and Its Implications for Structure Growth}},\ }\href {https://doi.org/10.3847/1538-4357/acfe06} {\bibfield  {journal} {\bibinfo  {journal} {Astrophys. J.}\ }\textbf {\bibinfo {volume} {962}},\ \bibinfo {pages} {112} (\bibinfo {year} {2024})},\ \Eprint {https://arxiv.org/abs/2304.05202} {arXiv:2304.05202 [astro-ph.CO]} \BibitemShut {NoStop}%
\bibitem [{\citenamefont {Madhavacheril}\ \emph {et~al.}(2024)\citenamefont {Madhavacheril} \emph {et~al.}}]{ACT:2023kun}%
  \BibitemOpen
  \bibfield  {author} {\bibinfo {author} {\bibfnamefont {M.~S.}\ \bibnamefont {Madhavacheril}} \emph {et~al.} (\bibinfo {collaboration} {ACT}),\ }\bibfield  {title} {\bibinfo {title} {{The Atacama Cosmology Telescope: DR6 Gravitational Lensing Map and Cosmological Parameters}},\ }\href {https://doi.org/10.3847/1538-4357/acff5f} {\bibfield  {journal} {\bibinfo  {journal} {Astrophys. J.}\ }\textbf {\bibinfo {volume} {962}},\ \bibinfo {pages} {113} (\bibinfo {year} {2024})},\ \Eprint {https://arxiv.org/abs/2304.05203} {arXiv:2304.05203 [astro-ph.CO]} \BibitemShut {NoStop}%
\bibitem [{\citenamefont {Brout}\ \emph {et~al.}(2022)\citenamefont {Brout} \emph {et~al.}}]{Brout:2022vxf}%
  \BibitemOpen
  \bibfield  {author} {\bibinfo {author} {\bibfnamefont {D.}~\bibnamefont {Brout}} \emph {et~al.},\ }\bibfield  {title} {\bibinfo {title} {{The Pantheon+ Analysis: Cosmological Constraints}},\ }\href {https://doi.org/10.3847/1538-4357/ac8e04} {\bibfield  {journal} {\bibinfo  {journal} {Astrophys. J.}\ }\textbf {\bibinfo {volume} {938}},\ \bibinfo {pages} {110} (\bibinfo {year} {2022})},\ \Eprint {https://arxiv.org/abs/2202.04077} {arXiv:2202.04077 [astro-ph.CO]} \BibitemShut {NoStop}%
\bibitem [{\citenamefont {Riess}\ \emph {et~al.}(2022)\citenamefont {Riess} \emph {et~al.}}]{Riess:2021jrx}%
  \BibitemOpen
  \bibfield  {author} {\bibinfo {author} {\bibfnamefont {A.~G.}\ \bibnamefont {Riess}} \emph {et~al.},\ }\bibfield  {title} {\bibinfo {title} {{A Comprehensive Measurement of the Local Value of the Hubble Constant with 1 km s$^{-1}$ Mpc$^{-1}$ Uncertainty from the Hubble Space Telescope and the SH0ES Team}},\ }\href {https://doi.org/10.3847/2041-8213/ac5c5b} {\bibfield  {journal} {\bibinfo  {journal} {Astrophys. J. Lett.}\ }\textbf {\bibinfo {volume} {934}},\ \bibinfo {pages} {L7} (\bibinfo {year} {2022})},\ \Eprint {https://arxiv.org/abs/2112.04510} {arXiv:2112.04510 [astro-ph.CO]} \BibitemShut {NoStop}%
\bibitem [{\citenamefont {Abbott}\ \emph {et~al.}(2024)\citenamefont {Abbott} \emph {et~al.}}]{DES:2024tys}%
  \BibitemOpen
  \bibfield  {author} {\bibinfo {author} {\bibfnamefont {T.~M.~C.}\ \bibnamefont {Abbott}} \emph {et~al.} (\bibinfo {collaboration} {DES}),\ }\href@noop {} {\bibinfo {title} {{The Dark Energy Survey: Cosmology Results With \textasciitilde{}1500 New High-redshift Type Ia Supernovae Using The Full 5-year Dataset}}} (\bibinfo {year} {2024}),\ \Eprint {https://arxiv.org/abs/2401.02929} {arXiv:2401.02929 [astro-ph.CO]} \BibitemShut {NoStop}%
\bibitem [{\citenamefont {Vincenzi}\ \emph {et~al.}(2024)\citenamefont {Vincenzi} \emph {et~al.}}]{DES:2024hip}%
  \BibitemOpen
  \bibfield  {author} {\bibinfo {author} {\bibfnamefont {M.}~\bibnamefont {Vincenzi}} \emph {et~al.} (\bibinfo {collaboration} {DES}),\ }\href@noop {} {\bibinfo {title} {{The Dark Energy Survey Supernova Program: Cosmological Analysis and Systematic Uncertainties}}} (\bibinfo {year} {2024}),\ \Eprint {https://arxiv.org/abs/2401.02945} {arXiv:2401.02945 [astro-ph.CO]} \BibitemShut {NoStop}%
\bibitem [{\citenamefont {Blas}\ \emph {et~al.}(2011)\citenamefont {Blas}, \citenamefont {Lesgourgues},\ and\ \citenamefont {Tram}}]{Diego_Blas_2011}%
  \BibitemOpen
  \bibfield  {author} {\bibinfo {author} {\bibfnamefont {D.}~\bibnamefont {Blas}}, \bibinfo {author} {\bibfnamefont {J.}~\bibnamefont {Lesgourgues}},\ and\ \bibinfo {author} {\bibfnamefont {T.}~\bibnamefont {Tram}},\ }\bibfield  {title} {\bibinfo {title} {The cosmic linear anisotropy solving system (class). part ii: Approximation schemes},\ }\href {https://doi.org/10.1088/1475-7516/2011/07/034} {\bibfield  {journal} {\bibinfo  {journal} {Journal of Cosmology and Astroparticle Physics}\ }\textbf {\bibinfo {volume} {2011}}\bibinfo  {number} { (07)},\ \bibinfo {pages} {034–034}}\BibitemShut {NoStop}%
\bibitem [{\citenamefont {Zuntz}\ \emph {et~al.}(2015)\citenamefont {Zuntz}, \citenamefont {Paterno}, \citenamefont {Jennings}, \citenamefont {Rudd}, \citenamefont {Manzotti}, \citenamefont {Dodelson}, \citenamefont {Bridle}, \citenamefont {Sehrish},\ and\ \citenamefont {Kowalkowski}}]{Zuntz:2014csq}%
  \BibitemOpen
\bibfield  {number} {  }\bibfield  {author} {\bibinfo {author} {\bibfnamefont {J.}~\bibnamefont {Zuntz}}, \bibinfo {author} {\bibfnamefont {M.}~\bibnamefont {Paterno}}, \bibinfo {author} {\bibfnamefont {E.}~\bibnamefont {Jennings}}, \bibinfo {author} {\bibfnamefont {D.}~\bibnamefont {Rudd}}, \bibinfo {author} {\bibfnamefont {A.}~\bibnamefont {Manzotti}}, \bibinfo {author} {\bibfnamefont {S.}~\bibnamefont {Dodelson}}, \bibinfo {author} {\bibfnamefont {S.}~\bibnamefont {Bridle}}, \bibinfo {author} {\bibfnamefont {S.}~\bibnamefont {Sehrish}},\ and\ \bibinfo {author} {\bibfnamefont {J.}~\bibnamefont {Kowalkowski}},\ }\bibfield  {title} {\bibinfo {title} {{CosmoSIS: modular cosmological parameter estimation}},\ }\href {https://doi.org/10.1016/j.ascom.2015.05.005} {\bibfield  {journal} {\bibinfo  {journal} {Astron. Comput.}\ }\textbf {\bibinfo {volume} {12}},\ \bibinfo {pages} {45} (\bibinfo {year} {2015})},\ \Eprint {https://arxiv.org/abs/1409.3409} {arXiv:1409.3409 [astro-ph.CO]} \BibitemShut {NoStop}%
\bibitem [{\citenamefont {Ballesteros}\ and\ \citenamefont {Lesgourgues}(2010)}]{Ballesteros:2010ks}%
  \BibitemOpen
  \bibfield  {author} {\bibinfo {author} {\bibfnamefont {G.}~\bibnamefont {Ballesteros}}\ and\ \bibinfo {author} {\bibfnamefont {J.}~\bibnamefont {Lesgourgues}},\ }\bibfield  {title} {\bibinfo {title} {{Dark energy with non-adiabatic sound speed: initial conditions and detectability}},\ }\href {https://doi.org/10.1088/1475-7516/2010/10/014} {\bibfield  {journal} {\bibinfo  {journal} {JCAP}\ }\textbf {\bibinfo {volume} {10}},\ \bibinfo {pages} {014}},\ \Eprint {https://arxiv.org/abs/1004.5509} {arXiv:1004.5509 [astro-ph.CO]} \BibitemShut {NoStop}%
\bibitem [{\citenamefont {Fang}\ \emph {et~al.}(2008)\citenamefont {Fang}, \citenamefont {Hu},\ and\ \citenamefont {Lewis}}]{Fang:2008sn}%
  \BibitemOpen
  \bibfield  {author} {\bibinfo {author} {\bibfnamefont {W.}~\bibnamefont {Fang}}, \bibinfo {author} {\bibfnamefont {W.}~\bibnamefont {Hu}},\ and\ \bibinfo {author} {\bibfnamefont {A.}~\bibnamefont {Lewis}},\ }\bibfield  {title} {\bibinfo {title} {{Crossing the Phantom Divide with Parameterized Post-Friedmann Dark Energy}},\ }\href {https://doi.org/10.1103/PhysRevD.78.087303} {\bibfield  {journal} {\bibinfo  {journal} {Phys. Rev. D}\ }\textbf {\bibinfo {volume} {78}},\ \bibinfo {pages} {087303} (\bibinfo {year} {2008})},\ \Eprint {https://arxiv.org/abs/0808.3125} {arXiv:0808.3125 [astro-ph]} \BibitemShut {NoStop}%
\bibitem [{\citenamefont {Crittenden}\ \emph {et~al.}(2012)\citenamefont {Crittenden}, \citenamefont {Zhao}, \citenamefont {Pogosian}, \citenamefont {Samushia},\ and\ \citenamefont {Zhang}}]{Crittenden:2011aa}%
  \BibitemOpen
  \bibfield  {author} {\bibinfo {author} {\bibfnamefont {R.~G.}\ \bibnamefont {Crittenden}}, \bibinfo {author} {\bibfnamefont {G.-B.}\ \bibnamefont {Zhao}}, \bibinfo {author} {\bibfnamefont {L.}~\bibnamefont {Pogosian}}, \bibinfo {author} {\bibfnamefont {L.}~\bibnamefont {Samushia}},\ and\ \bibinfo {author} {\bibfnamefont {X.}~\bibnamefont {Zhang}},\ }\bibfield  {title} {\bibinfo {title} {{Fables of reconstruction: controlling bias in the dark energy equation of state}},\ }\href {https://doi.org/10.1088/1475-7516/2012/02/048} {\bibfield  {journal} {\bibinfo  {journal} {JCAP}\ }\textbf {\bibinfo {volume} {02}},\ \bibinfo {pages} {048}},\ \Eprint {https://arxiv.org/abs/1112.1693} {arXiv:1112.1693 [astro-ph.CO]} \BibitemShut {NoStop}%
\bibitem [{\citenamefont {Gilks}\ \emph {et~al.}(1995)\citenamefont {Gilks}, \citenamefont {Richardson},\ and\ \citenamefont {Spiegelhalter}}]{gilks:1995}%
  \BibitemOpen
  \bibfield  {author} {\bibinfo {author} {\bibfnamefont {W.}~\bibnamefont {Gilks}}, \bibinfo {author} {\bibfnamefont {S.}~\bibnamefont {Richardson}},\ and\ \bibinfo {author} {\bibfnamefont {D.}~\bibnamefont {Spiegelhalter}},\ }\href {https://books.google.it/books?id=TRXrMWY\_i2IC} {\emph {\bibinfo {title} {Markov Chain Monte Carlo in Practice}}},\ Chapman \& Hall/CRC Interdisciplinary Statistics\ (\bibinfo  {publisher} {Taylor \& Francis},\ \bibinfo {year} {1995})\BibitemShut {NoStop}%
\bibitem [{\citenamefont {{Goodman}}\ and\ \citenamefont {{Weare}}(2010)}]{aff_inv}%
  \BibitemOpen
  \bibfield  {author} {\bibinfo {author} {\bibfnamefont {J.}~\bibnamefont {{Goodman}}}\ and\ \bibinfo {author} {\bibfnamefont {J.}~\bibnamefont {{Weare}}},\ }\bibfield  {title} {\bibinfo {title} {{Ensemble samplers with affine invariance}},\ }\href {https://doi.org/10.2140/camcos.2010.5.65} {\bibfield  {journal} {\bibinfo  {journal} {Communications in Applied Mathematics and Computational Science}\ }\textbf {\bibinfo {volume} {5}},\ \bibinfo {pages} {65} (\bibinfo {year} {2010})}\BibitemShut {NoStop}%
\bibitem [{\citenamefont {Foreman-Mackey}\ \emph {et~al.}(2013)\citenamefont {Foreman-Mackey}, \citenamefont {Hogg}, \citenamefont {Lang},\ and\ \citenamefont {Goodman}}]{Foreman_Mackey_2013}%
  \BibitemOpen
  \bibfield  {author} {\bibinfo {author} {\bibfnamefont {D.}~\bibnamefont {Foreman-Mackey}}, \bibinfo {author} {\bibfnamefont {D.~W.}\ \bibnamefont {Hogg}}, \bibinfo {author} {\bibfnamefont {D.}~\bibnamefont {Lang}},\ and\ \bibinfo {author} {\bibfnamefont {J.}~\bibnamefont {Goodman}},\ }\bibfield  {title} {\bibinfo {title} {\texttt{emcee}: The mcmc hammer},\ }\href {https://doi.org/10.1086/670067} {\bibfield  {journal} {\bibinfo  {journal} {Publications of the Astronomical Society of the Pacific}\ }\textbf {\bibinfo {volume} {125}},\ \bibinfo {pages} {306–312} (\bibinfo {year} {2013})}\BibitemShut {NoStop}%
\bibitem [{\citenamefont {Metropolis}\ \emph {et~al.}(1953)\citenamefont {Metropolis}, \citenamefont {Rosenbluth}, \citenamefont {Rosenbluth}, \citenamefont {Teller},\ and\ \citenamefont {Teller}}]{Metropolis:1953am}%
  \BibitemOpen
  \bibfield  {author} {\bibinfo {author} {\bibfnamefont {N.}~\bibnamefont {Metropolis}}, \bibinfo {author} {\bibfnamefont {A.~W.}\ \bibnamefont {Rosenbluth}}, \bibinfo {author} {\bibfnamefont {M.~N.}\ \bibnamefont {Rosenbluth}}, \bibinfo {author} {\bibfnamefont {A.~H.}\ \bibnamefont {Teller}},\ and\ \bibinfo {author} {\bibfnamefont {E.}~\bibnamefont {Teller}},\ }\bibfield  {title} {\bibinfo {title} {{Equation of state calculations by fast computing machines}},\ }\href {https://doi.org/10.1063/1.1699114} {\bibfield  {journal} {\bibinfo  {journal} {J. Chem. Phys.}\ }\textbf {\bibinfo {volume} {21}},\ \bibinfo {pages} {1087} (\bibinfo {year} {1953})}\BibitemShut {NoStop}%
\bibitem [{\citenamefont {Hastings}(1970)}]{Hastings:1970aa}%
  \BibitemOpen
  \bibfield  {author} {\bibinfo {author} {\bibfnamefont {W.~K.}\ \bibnamefont {Hastings}},\ }\bibfield  {title} {\bibinfo {title} {{Monte Carlo Sampling Methods Using Markov Chains and Their Applications}},\ }\href {https://doi.org/10.1093/biomet/57.1.97} {\bibfield  {journal} {\bibinfo  {journal} {Biometrika}\ }\textbf {\bibinfo {volume} {57}},\ \bibinfo {pages} {97} (\bibinfo {year} {1970})}\BibitemShut {NoStop}%
\bibitem [{\citenamefont {Lewis}(2019)}]{Lewis:2019xzd}%
  \BibitemOpen
  \bibfield  {author} {\bibinfo {author} {\bibfnamefont {A.}~\bibnamefont {Lewis}},\ }\href {https://getdist.readthedocs.io} {\bibinfo {title} {{GetDist: a Python package for analysing Monte Carlo samples}}} (\bibinfo {year} {2019}),\ \Eprint {https://arxiv.org/abs/1910.13970} {arXiv:1910.13970 [astro-ph.IM]} \BibitemShut {NoStop}%
\bibitem [{\citenamefont {Nelder}\ and\ \citenamefont {Mead}(1965)}]{Nelder:1965zz}%
  \BibitemOpen
  \bibfield  {author} {\bibinfo {author} {\bibfnamefont {J.~A.}\ \bibnamefont {Nelder}}\ and\ \bibinfo {author} {\bibfnamefont {R.}~\bibnamefont {Mead}},\ }\bibfield  {title} {\bibinfo {title} {{A Simplex Method for Function Minimization}},\ }\href {https://doi.org/10.1093/comjnl/7.4.308} {\bibfield  {journal} {\bibinfo  {journal} {Comput. J.}\ }\textbf {\bibinfo {volume} {7}},\ \bibinfo {pages} {308} (\bibinfo {year} {1965})}\BibitemShut {NoStop}%
\bibitem [{\citenamefont {Efstathiou}(2024)}]{Efstathiou:2024xcq}%
  \BibitemOpen
  \bibfield  {author} {\bibinfo {author} {\bibfnamefont {G.}~\bibnamefont {Efstathiou}},\ }\href@noop {} {\bibinfo {title} {{Evolving Dark Energy or Supernovae Systematics?}}} (\bibinfo {year} {2024}),\ \Eprint {https://arxiv.org/abs/2408.07175} {arXiv:2408.07175 [astro-ph.CO]} \BibitemShut {NoStop}%
\bibitem [{\citenamefont {Planck Collaboration~V}(2020)}]{planck:2018like}%
  \BibitemOpen
  \bibfield  {author} {\bibinfo {author} {\bibfnamefont {.}~\bibnamefont {Planck Collaboration~V}} (\bibinfo {collaboration} {Planck}),\ }\bibfield  {title} {\bibinfo {title} {V - cmb power spectra and likelihoods},\ }\href {https://doi.org/10.1051/0004-6361/201936386} {\bibfield  {journal} {\bibinfo  {journal} {Astron. Astrophys.}\ }\textbf {\bibinfo {volume} {641}},\ \bibinfo {pages} {A5} (\bibinfo {year} {2020})},\ \Eprint {https://arxiv.org/abs/1907.12875} {arXiv:1907.12875 [astro-ph.CO]} \BibitemShut {NoStop}%
\bibitem [{\citenamefont {Planck Collaboration~III}(2020)}]{planck:2018_maps}%
  \BibitemOpen
  \bibfield  {author} {\bibinfo {author} {\bibfnamefont {.}~\bibnamefont {Planck Collaboration~III}} (\bibinfo {collaboration} {Planck}),\ }\bibfield  {title} {\bibinfo {title} {{III - High Frequency Instrument data processing and frequency maps}},\ }\href {https://doi.org/10.1051/0004-6361/201832909} {\bibfield  {journal} {\bibinfo  {journal} {Astron. Astrophys.}\ }\textbf {\bibinfo {volume} {641}},\ \bibinfo {pages} {A3} (\bibinfo {year} {2020})},\ \Eprint {https://arxiv.org/abs/1807.06207} {arXiv:1807.06207 [astro-ph.CO]} \BibitemShut {NoStop}%
\bibitem [{\citenamefont {Vincenzi}\ \emph {et~al.}(2025)\citenamefont {Vincenzi} \emph {et~al.}}]{DES:2025tir}%
  \BibitemOpen
  \bibfield  {author} {\bibinfo {author} {\bibfnamefont {M.}~\bibnamefont {Vincenzi}} \emph {et~al.} (\bibinfo {collaboration} {DES}),\ }\href@noop {} {\bibinfo {title} {{Comparing the DES-SN5YR and Pantheon+ SN cosmology analyses: Investigation based on ''Evolving Dark Energy or Supernovae systematics?''}}} (\bibinfo {year} {2025}),\ \Eprint {https://arxiv.org/abs/2501.06664} {arXiv:2501.06664 [astro-ph.CO]} \BibitemShut {NoStop}%
\bibitem [{\citenamefont {Dhawan}\ \emph {et~al.}(2024)\citenamefont {Dhawan}, \citenamefont {Popovic},\ and\ \citenamefont {Goobar}}]{Dhawan:2024gqy}%
  \BibitemOpen
  \bibfield  {author} {\bibinfo {author} {\bibfnamefont {S.}~\bibnamefont {Dhawan}}, \bibinfo {author} {\bibfnamefont {B.}~\bibnamefont {Popovic}},\ and\ \bibinfo {author} {\bibfnamefont {A.}~\bibnamefont {Goobar}},\ }\href@noop {} {\bibinfo {title} {{The axis of systematic bias in SN\textasciitilde{}Ia cosmology and implications for DESI 2024 results}}} (\bibinfo {year} {2024}),\ \Eprint {https://arxiv.org/abs/2409.18668} {arXiv:2409.18668 [astro-ph.CO]} \BibitemShut {NoStop}%
\bibitem [{\citenamefont {Notari}\ \emph {et~al.}(2024)\citenamefont {Notari}, \citenamefont {Redi},\ and\ \citenamefont {Tesi}}]{Notari:2024zmi}%
  \BibitemOpen
  \bibfield  {author} {\bibinfo {author} {\bibfnamefont {A.}~\bibnamefont {Notari}}, \bibinfo {author} {\bibfnamefont {M.}~\bibnamefont {Redi}},\ and\ \bibinfo {author} {\bibfnamefont {A.}~\bibnamefont {Tesi}},\ }\href@noop {} {\bibinfo {title} {{BAO vs. SN evidence for evolving dark energy}}} (\bibinfo {year} {2024}),\ \Eprint {https://arxiv.org/abs/2411.11685} {arXiv:2411.11685 [astro-ph.CO]} \BibitemShut {NoStop}%
\bibitem [{\citenamefont {Colg\'ain}\ and\ \citenamefont {Sheikh-Jabbari}(2024)}]{Colgain:2024mtg}%
  \BibitemOpen
  \bibfield  {author} {\bibinfo {author} {\bibfnamefont {E.~O.}\ \bibnamefont {Colg\'ain}}\ and\ \bibinfo {author} {\bibfnamefont {M.~M.}\ \bibnamefont {Sheikh-Jabbari}},\ }\href@noop {} {\bibinfo {title} {{DESI and SNe: Dynamical Dark Energy, $\Omega_m$ Tension or Systematics?}}} (\bibinfo {year} {2024}),\ \Eprint {https://arxiv.org/abs/2412.12905} {arXiv:2412.12905 [astro-ph.CO]} \BibitemShut {NoStop}%
\bibitem [{\citenamefont {Huang}\ \emph {et~al.}(2025)\citenamefont {Huang}, \citenamefont {Cai},\ and\ \citenamefont {Wang}}]{Huang:2025som}%
  \BibitemOpen
  \bibfield  {author} {\bibinfo {author} {\bibfnamefont {L.}~\bibnamefont {Huang}}, \bibinfo {author} {\bibfnamefont {R.-G.}\ \bibnamefont {Cai}},\ and\ \bibinfo {author} {\bibfnamefont {S.-J.}\ \bibnamefont {Wang}},\ }\href@noop {} {\bibinfo {title} {{The DESI 2024 hint for dynamical dark energy is biased by low-redshift supernovae}}} (\bibinfo {year} {2025}),\ \Eprint {https://arxiv.org/abs/2502.04212} {arXiv:2502.04212 [astro-ph.CO]} \BibitemShut {NoStop}%
\bibitem [{\citenamefont {Zumalacarregui}(2020)}]{Zumalacarregui:2020cjh}%
  \BibitemOpen
  \bibfield  {author} {\bibinfo {author} {\bibfnamefont {M.}~\bibnamefont {Zumalacarregui}},\ }\bibfield  {title} {\bibinfo {title} {{Gravity in the Era of Equality: Towards solutions to the Hubble problem without fine-tuned initial conditions}},\ }\href {https://doi.org/10.1103/PhysRevD.102.023523} {\bibfield  {journal} {\bibinfo  {journal} {Phys. Rev. D}\ }\textbf {\bibinfo {volume} {102}},\ \bibinfo {pages} {023523} (\bibinfo {year} {2020})},\ \Eprint {https://arxiv.org/abs/2003.06396} {arXiv:2003.06396 [astro-ph.CO]} \BibitemShut {NoStop}%
\bibitem [{\citenamefont {Chudaykin}\ and\ \citenamefont {Kunz}(2024)}]{Chudaykin:2024gol}%
  \BibitemOpen
  \bibfield  {author} {\bibinfo {author} {\bibfnamefont {A.}~\bibnamefont {Chudaykin}}\ and\ \bibinfo {author} {\bibfnamefont {M.}~\bibnamefont {Kunz}},\ }\bibfield  {title} {\bibinfo {title} {{Modified gravity interpretation of the evolving dark energy in light of DESI data}},\ }\href {https://doi.org/10.1103/PhysRevD.110.123524} {\bibfield  {journal} {\bibinfo  {journal} {Phys. Rev. D}\ }\textbf {\bibinfo {volume} {110}},\ \bibinfo {pages} {123524} (\bibinfo {year} {2024})},\ \Eprint {https://arxiv.org/abs/2407.02558} {arXiv:2407.02558 [astro-ph.CO]} \BibitemShut {NoStop}%
\end{thebibliography}%

\end{document}